\documentclass[twocolumn,amsmath,amssymb,aps,preprintnumbers,prd]{revtex4-1}
\usepackage{graphicx} 
\usepackage{xcolor}
\usepackage{slashed} 
\usepackage{hyperref}
\usepackage{bm}
\usepackage{soul} 
\allowdisplaybreaks
\def\beq{\begin{equation}}
\def\eeq{\end{equation}}
\def\beqn{\begin{eqnarray}} 
\def\eeqn{\end{eqnarray}}
\def\nn{\nonumber}

\newcommand{\valencia}{Instituto de F\'isica Corpuscular, Universitat de Val\`{e}ncia -- 
Consejo Superior de Investigaciones Cient\'{\i}ficas, Parc Cient\'{\i}fic, 46980 Paterna, Valencia, Spain.}
\newcommand{\regensburg}{Institut f\"ur Theoretische Physik, Universit\"at Regensburg, 93040 Regensburg, Germany.}
\newcommand{\madrid}{Departamento de F\'isica Te\'orica \& IPARCOS, Universidad Complutense de Madrid, E-28040 Madrid, Spain}
\newcommand{\salamanca}{Departamento de F\'isica Fundamental e IUFFyM, Universidad de Salamanca, 37008 Salamanca, Spain.}
\newcommand{\culiacan}{Facultad de Ciencias F\'isico-Matem\'aticas, Universidad Aut\'onoma de Sinaloa, Ciudad Universitaria, CP 80000 Culiac\'an, Sinaloa, M\'exico.}

\begin{document}
\title{Using analytic models to describe effective PDFs}
\author{
S. A. Ochoa-Oregon$^{a}$,
D. F. Renter\'ia-Estrada$^{b}$,
R. J. Hern\'andez-Pinto$^{a}$,
G. F. R. Sborlini$^{c}$,
P. Zurita$^{d,e}$
}
\affiliation{
$^{a}$ \culiacan\\
$^{b}$ \valencia\\
$^{c}$ \salamanca\\
$^{d}$ \regensburg\\
$^{e}$ \madrid}

\preprint{IPARCOS-UCM-24-023}

\begin{abstract}
{Parton distribution functions play a pivotal role in hadron collider phenomenology. They are non-perturbative quantities extracted from fits to available data, and their scale dependence is dictated by the DGLAP evolution equations. In this article, we discuss {machine-assisted} strategies to efficiently compute PDFs {directly incorporating the scale evolution without the need of separately solving DGLAP equations.} Analytical approximations to the PDFs as a function of $x$ and $Q^2$, including up to next-to-leading order effects in Quantum Chromodynamics, are obtained. The methodology is tested by reproducing the \texttt{HERAPDF2.0} set and implementing the analytical expressions in benchmarking codes. It is found that the computational cost is reduced while the precision of the simulations stays well under control.} 
\end{abstract}


\setcounter{page}{1}
\maketitle

\section{Introduction}
\label{sec:Introduction}
Breaking the precision frontier in particle physics is a challenging task. The tiny discrepancies among experiments and theoretical predictions might hide new phenomena, and this forces theoreticians to refine as much as possible their methodologies to produce accurate simulations of particle collisions. Even if a plethora of powerful methods are available (see e.g. Ref. \cite{Heinrich:2020ybq}), most of them require enormous computing resources to achieve the intended precision. In this direction, the purpose of this investigation is to reduce the resource consumption of higher-order calculations for perturbative QFT whilst keeping the precision achieved by the most advanced theoretical results. This will allow to provide fast and reliable results, and also to reduce the environmental impact associated to the high-energy research \cite{Banerjee:2023avd}. 

Most of the current higher-order cross-section simulations rely on the factorization theorem \cite{Collins:1989gx,Catani:2011st,Cieri:2024ytf}, i.e., splitting the whole calculation into perturbative and non-perturbative sectors. The process-dependent perturbative contribution is given in terms of well-defined analytical functions (or highly-efficient numerical representations of them). The universal, non-perturbative part, encoded in parton distribution functions (PDFs), can not be computed from first principles: only their evolution with the factorization scale chosen can be calculated via the DGLAP \cite{Altarelli:1977zs} evolution equations. The determination of the PDFs is, instead, achieved by performing global fits to existing data. For practical implementations, the outcome of these fits are given as tables in the relevant kinematic variables with a function that interpolates over them. The interpolation requires to perform several evaluations and the routines are written to be as fast as possible to allow for the efficient computation of physical observables. However, some cross-sections might require a very large number of trials (and running time) to achieve relevant precision, due to the nature of the process. Thus, it is worth exploring the possibility of reducing the CPU/GPU time required by providing an alternative form for the PDFs. Therefore, in this work, we aim to determine an analytical functional form for a set of known PDFs, by approximating their $(x,Q^2)$-behaviour. This is equivalent to identify approximations to closed analytic solutions of DGLAP equations, a highly non-trivial problem in the context of coupled integro-differential equations \cite{Simonelli:2024vyh}.

To achieve this ambitious objective, {we profit from artificial intelligence, machine-learning and similar tools implemented within \texttt{Mathematica} built-in functions. By using machine-assisted techniques, we identify suitable functional dependencies to accurately describe the PDF sets.} Our thesis is that, avoiding the interpolation over the grids, it will drastically reduce the time required to perform the simulations/computations. Furthermore, the PDFs would be then written in terms of a few parameters, reducing the storage required for them. Also, having access to analytic formulae for PDFs will allow to calculate their derivatives, both in $x$ and $Q$, which are relevant for performing the matching of $F_2$ in the threshold regions.

Throughout this work we focus on the set of collinear proton PDFs extracted by the \texttt{xFitter} collaboration, \texttt{HERAPDF2.0} \cite{H1:2015ubc}. { We would like to highlight that our methodology is applicable to any PDF set, independently of the order at which it is calculated: if the PDF set exists, it is possible to find an analytic approximation with the techniques explained in this article. For this reason, }we stress that our study does not replace, in any way, the need of performing global fits. On the contrary, it is only by having the PDF sets already determined that we can search a posteriori for a functional form. {Our analytic approximations to a PDF set constitute a first proof-of-concept, and we show that it successfully passed several physically-motivated quality checks.}

This work is organised as follows. In Sec. \ref{sec:Thesis} we present our proposed analytical model for a set of collinear, unpolarised proton PDFs, using \texttt{HERAPDF2.0} as baseline. The methodology used to determine the functional form is explained in detail in Sec. \ref{sec:Methodology}. Our results and their comparison with the baseline PDFs can be found in Sec. \ref{sec:Results}. {In Sec. \ref{sec:Comparison}, we carefully quantify the errors induced by our analytic approximations and check the validity of the sum rules. There, we also check} the time required to run benchmarking codes using both the available PDF grids and our a posteriori analytical PDFs. We conclude by summarising our findings in Sec. \ref{sec:Conclusions}.


\section{Starting hypothesis}
\label{sec:Thesis}
Traditionally, proton PDFs \footnote{This also applies, with some differences, to other non-perturbative quantities feasible of phenomenological studies in perturbative Quantum Chromodynamics (pQCD), such as fragmentation functions, fracture functions, transverse momentum dependent distributions, etc.} are determined by proposing a functional form in Bjorken $x$  for the parton densities (or a linear combination of them) at some chosen initial scale $Q_0^2$. Giving values to the parameters involved and evolving the PDFs to the experimental scales $Q^2$ using the DGLAP equations, one obtains the distributions that, convoluted with the partonic cross-sections, is compared with data. Repeating the procedure until an adequate description is achieved (according to some chosen criteria), the best fit parameters are determined. This is the case of many sets of proton and nuclear PDFs (e.g. \cite{Bailey:2020ooq,Hou:2019efy,Alekhin:2017kpj,Helenius:2021tof,deFlorian:2011fp,Eskola:2021nhw,Duwentaster:2021ioo}), and fragmentation functions (e.g.\cite{Borsa:2022vvp,Borsa:2023zxk}). Other possibility that does not use a proposed parametrization but rather relies on Neural Networks is also employed by the NNPDF collaboration (see e.g. Ref. \cite{NNPDF:2021njg} and references within). 

Regarding the parametric form of the distributions, they are usually chosen to be an Euler beta function, with some extra flexibility given by a multiplicative polynomial or exponential. Schematically: 
\begin{equation}
f_{i}(x,Q_0^{2})=N_i x^{\alpha_i} (1-x)^{\beta_{i}} P(x,c_{ij})\, ,
\label{eq:basic_pdf}
\end{equation}
where $x$ is, in the Breit frame, the fraction of momentum of the proton carried by the parton, $i$ indicates a parton flavour or a combination of them (selected to do the evolution), and $P(x,c_{ij})$ is some function of $x$ with coefficients $c_{ij}$. This form is flexible enough that very different shapes can be achieved by varying the parameters. After the best fit is found, tables in $x$ and $Q^2$ are made available (in modern times through \texttt{LHAPDF} \cite{Butterworth:2015oua,Buckley:2014ana}), with fast interpolating routines. 

To determine an analytical expression for the collinear proton PDFs, which we test with the \texttt{HERAPDF2.0} set at next-to-leading order (NLO) accuracy in QCD, we propose that the $Q^2$ dependence of the PDFs is given by an extension of Eq. (\ref{eq:basic_pdf}), with each parameter acquiring a $Q^2$ dependence. In other words, we start from the assumption that
\begin{align}
\nn f_{i}(x,Q^{2})&=N_i(Q^2) x^{\alpha_i(Q^2)} \,  (1-x)^{\beta_{i}(Q^2)}  
\\ & \times P(x,c_{ij}(Q^2)) \, ,
\label{eq:fit_xPDF(X,Q)}
\end{align}
with $P$ being, e.g., a polynomial in $x$. The aim of our work is to determine these functions using ML methods, and explore up to what extent the $Q^2$ dependence can be reproduced with these functional forms.


\section{Methodology}
\label{sec:Methodology}
In order to obtain the coefficients from Eq. (\ref{eq:fit_xPDF(X,Q)}), we performed a two-step fitting procedure. First, we fixed a PDF set within the \texttt{LHAPDF} framework: in our study, it was \texttt{HERAPDF20$\_$NLO$\_$EIG}, obtained by the \texttt{xFitter} group, which uses solely data from the HERA collider. {Then, we generated a grid of ${\cal O}(5000)$ random points in $\{x,Q\}$ with $x \in [10^{-4},0.65]$ and $Q \in [Q_{\rm Min},1000]$, evaluating the corresponding PDF in these points.} The range in $x$ was chosen in accordance to the recommendations done in Ref. \cite{H1:2015ubc}. For the gluon and light quarks, we set $Q_{\rm Min}=Q_0\approx 1.37 $ GeV, whilst we used $Q_{\rm Min}= 1.5 $ GeV and $Q_{\rm Min}= 4.5 $ GeV for the charm and bottom quark, respectively.

Regarding the parametrization, we draw inspiration from Ref. \cite{H1:2015ubc}, where at the initial scale $Q_0$ the authors proposed
\begin{eqnarray}
x \, u_v(x) &=& A_{u_v} \, x^{B_{u_v}} (1-x)^{C_{u_v}} \, (1+E_{u_v}\, x^2)\, ,
\label{eq:xuvHERA}
\\ x \, d_v(x) &=& A_{d_v} \, x^{B_{d_v}} (1-x)^{C_{d_v}} \, ,
\label{eq:xdvHERA}
\\ x \, {\bar{U}}(x) &=& A_{\bar{U}} \, x^{B_{\bar{U}}} (1-x)^{C_{\bar{U}}} \, (1+D_{\bar{U}}\, x^2) \, ,
\label{eq:xUbarHERA}
\\ x \, {\bar{D}}(x) &=& A_{\bar{D}} \, x^{B_{\bar{D}}} (1-x)^{C_{\bar{D}}} \, ,
\label{eq:xDbarHERA}
\\ x\, \bar{s}(x) &=& f_s\, x\, \bar{D}(x) \, ,
\label{eq:xsHERA}
\end{eqnarray}
for the valence and sea distributions of light quarks ($u$, $d$ and $s$), and 
\begin{eqnarray}
x \, g(x) &=& A_g \, x^{B_g} (1-x)^{C_g} \, - A'_{g} \, x^{B'_g} (1-x)^{C'_g} \, ,
\label{eq:xgHERA}
\end{eqnarray}
for the gluon. In these formulae, we are using $f_s=0.4$ at $Q_0$ and $s \equiv \bar{s}$, as well as
\beq
\bar{U} = \bar{u} \, , \quad \bar{D} = \bar{d} + \bar{s} \, .
\eeq
When the scales probed are above the production threshold of a heavy-quark, said quark will be radiatively produced and thus have a nonzero distribution (unless intrinsic heavy flavours are considered, which is beyond the scope of the present work). In particular, in \texttt{HERAPDF2.0} the optimised general mass variable-flavour-number scheme RTOPT was used to treat charm and bottom PDFs \cite{Thorne:2012az}. Due to their radiative nature, the gluon has a noticeable impact in the determination of sea and heavy-quark distributions.

Motivated by this, we proposed functional forms like Eq. (\ref{eq:fit_xPDF(X,Q)}) to fit $u_v(x,Q^2)$ and $d_v(x,Q^2)$, whilst we relied on gluon-like sums of Euler beta functions, i.e. Eq. (\ref{eq:xgHERA}), for the remaining distributions. To perform a quantitative study of the goodness of the approximation, we used the so-called \emph{integral error}, defined according to
\beq
\Delta_i(Q^2) = \frac{I\left[f^{\rm ML}_i(x,Q^2),Q^2\right]-I\left[f^{\rm HERA}_i(x,Q^2),Q^2\right]}{I\left[f^{\rm HERA}_i(x,Q^2),Q^2\right]} \, ,
\label{eq:IntegralError}
\eeq
with $f_i^{\rm ML}$ and $f_i^{\rm HERA}$ being the PDF corresponding to the flavour $i$ obtained with our analytic approximation (ML-PDF from here onwards) and the \texttt{HERAPDF2.0} set, respectively. The integration operator $I\left[f,Q^2\right]$ is given by
\beq
I\left[f,Q^2\right] = \int_{10^{-4}}^1 \, dx \, f(x,Q^2) \, .
\eeq
We would like to highlight that this definition of the error is suitable to control the validity of the sum rules. Also, since the PDFs contribute to the hadronic cross-section through a convolution with the partonic cross-section, we observed that the integral error successfully provides an error estimation for the physical observables. { Both tests are discussed extensively in Sec. \ref{sec:Comparison}.}  

{ Besides this error definition, we explored additional estimators, such as the \emph{error shape:}
\beq
\tilde{\Delta}_i(Q^2) = \frac{1}{N}\, \sum_{j=1}^N \Bigg\vert 1-\frac{f^{\rm ML}_i(x_j,Q^2)}{f^{\rm HERA}_i(x_j,Q^2)}\Bigg\vert  \, ,
\label{eq:ShapeError}
\eeq
where $\{x_j\}_{j=1,\ldots,N}$ is a partition of $x \in [10^{-4},1]$. This definition is particularly sensitive to fluctuations or oscillations around the original PDF, and leads to a noticeable overestimation of the error. For example, we found typical error shapes of ${\cal O}(10-50 \, \%)$ for a large range of $Q$ values, in spite of percent and even sub-percent level deviations of the sum rules and other physical observables (see Sec. \ref{sec:Comparison}). For this reason, we consider the \emph{integral error} as a more reliable estimator.}


\section{Results}
\label{sec:Results}
In this section, we explicitly report the expressions for the analytic approximations to all the PDF flavours provided by \texttt{HERAPDF2.0} at NLO. {We fitted the $u_v$, $\bar{u}$, $d_v$, $\bar{d}$, $g$, $s$, $c$ and $b$ distributions following the functional forms proposed in each of the following sub-sections.} For each value of $Q$, we extracted the coefficients $\left\{A_i,B_i, \ldots\right\}$ performing a fit in $x$ with {the \texttt{Mathematica} built-in function \texttt{NonlinearModelFit} \cite{reference.wolfram_2023_nonlinearmodelfit}}. 

Then, we performed a second fit to determine the $Q$-dependence of these coefficients. {In this step, we relied on machine-assisted techniques that provide suitable functional forms to build the ansatzes. In particular, we make use of the \texttt{Mathematica} built-in routine \texttt{FindFormula} to generate an approximation to the coefficients $\left\{A_i,B_i, \ldots\right\}$ as functions of $Q$. Then, we used these approximations to propose an ansatz for each coefficient, including similar expressions to those found by \texttt{FindFormula}}. The functions obtained after this two-step fitting procedure, together with the full set of analytic ML-PDFs {are publicly available in an ancillary file uploaded to the Zenodo repository \cite{ZENODOPDF}}. Given that the final number of parameters is quite large, we refrain from presenting in the manuscript a table with the fitted parameters and functional forms.

\subsection{Gluon distribution}
\label{ssec:Gluon}
To obtain an approximation to the gluon distribution, we proposed a functional form inspired by Eq. (\ref{eq:xgHERA}). Concretely, we used
\beq
x \, g(x,Q^2) = f_1(x,Q^2) - \Theta(x_{C,g}-x)\, f_2(x,Q^2) \, ,
\label{eq:GenG}
\eeq
with $f_i$, for $i=1,2$, as given in Eq. (\ref{eq:basic_pdf}). In this way, for values of $x$ below the threshold $x_{C,g}$, Eq. (\ref{eq:GenG}) offers more flexibility to fit the PDF, in a fully analogous way as done in Ref. \cite{H1:2015ubc}.

\begin{figure}[htb]
\centering
\includegraphics[scale=0.5]{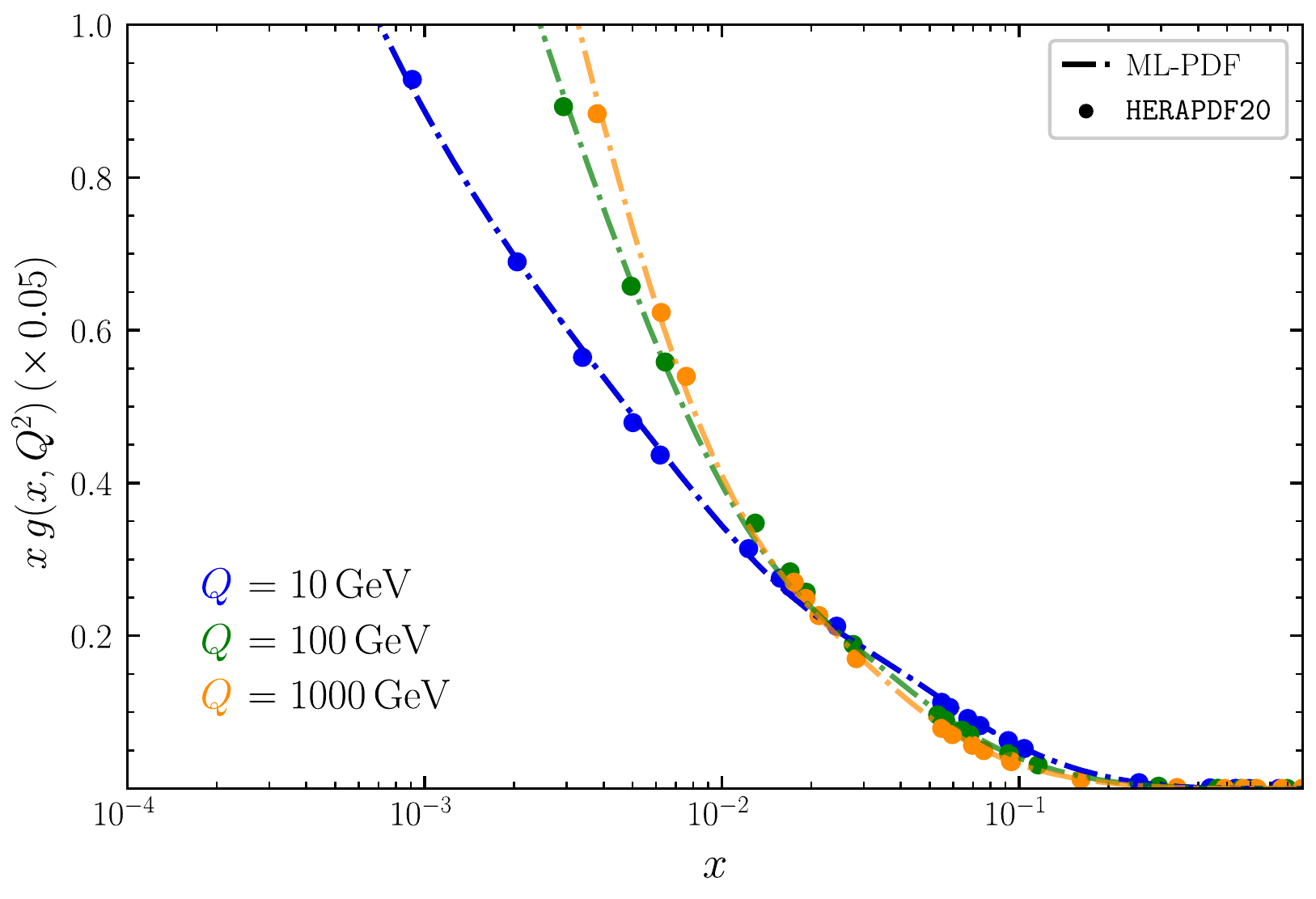}
\caption{Comparison between our analytic ML-PDF approximation (dashed lines) and \texttt{HERAPDF20} (solid dots) for gluon distributions, at three different values of $Q$. Notice that the gluon PDFs are multiplied by a factor $0.05$ in order to better appreciate the differences among them at different scales.}
\label{fig:Gdistribution}
\end{figure}

Furthermore, we noticed that the $Q$-dependence in the coefficient vary significantly in the whole $Q$ range considered (i.e. from $Q_0$ to 1000 GeV). Hence, we split the analysis in four separate regions:
\beqn
R_1 &=& \{Q_0 \leq Q \leq 2.5 \, {\rm GeV}\} \, ,
\\ R_2 &=& \{2.5 \, {\rm GeV} \leq Q \leq 5 \, {\rm GeV}\} \, ,
\\ R_3 &=& \{5 \, {\rm GeV} \leq Q \leq 150 \, {\rm GeV}\} \, ,
\\ R_4 &=& \{150 \, {\rm GeV} \leq Q \leq 1000 \, {\rm GeV}\} \, ,
\eeqn
which we determined by examining the $Q$ behaviour of the gluon density.
Consequently, we defined 
\beqn
\nn x \, g(x,Q^2) &=& A_{g}(Q^2) \, x^{B_{g}(Q^2)} \, (1-x)^{C_{g}(Q^2)} 
\\ \nn &-& \Theta(x_{C,g}-x) A'_{g}(Q^2) \, x^{B'_{g}(Q^2)} 
\\ &\times&  (1-x)^{C'_{g}(Q^2)} \, ,
\label{gR1}
\eeqn
for $Q \in R_1$ and 
\beqn
\nn x \, g(x,Q^2) &=& A_{g}(Q^2) \, x^{B_{g}(Q^2)} \, (1-x)^{C_{g}(Q^2)} 
\\ \nn &-& \Theta(x_{C,g}-x) A'_{g}(Q^2) \, x^{B'_{g}(Q^2)} 
\\ &\times& (1-x)^{C'_{g}(Q^2)}[1+D'_g(Q^2) x^2] \, ,
\label{gR234}
\eeqn
for $Q \in \{R_2,R_3,R_4\}$, together with $x_{C,g}=0.1$. Again, this last value was fixed by an exploratory procedure.

We emphasize that, since the analysis was done independently for each region, the functions $\{A_g(Q^2),B_g(Q^2),C_g(Q^2)\}$ and $\{A'_g(Q^2),B'_g(Q^2),C'_g(Q^2),D'_g(Q^2)\}$ in Eq. (\ref{gR234}) have different behaviours in $R_2$, $R_3$ and $R_4$.

In Fig. \ref{fig:Gdistribution}, we show our analytic ML-PDF approximation to $x\, g(x,Q^2)$ w.r.t. the corresponding gluon PDF from \texttt{HERAPDF20$\_$NLO$\_$EIG}, {at $10$ GeV (blue), $100$ GeV (green) and $1000$ GeV (orange)}. Even if small fluctuations occur at low-$x$, the overall agreement is very good. Furthermore, the agreement remains when choosing different values of $Q$, thanks to splitting the analysis into four regions.

\subsection{Down quark distributions}
\label{ssec:DOWN}
In this case, the optimal fit was achieved by considering $d_v$ and $\bar{d}$ distributions. Therefore, we wrote:
\beqn
x \, d_v(x,Q^2) &=& A_{d_v}(Q^2) \, x^{B_{d_v}(Q^2)} \, (1-x)^{C_{d_v}(Q^2)} 
\label{dVALENCIA}
\\ \nn &\times& \left[1+D_{d_v}(Q^2)x^2+E_{d_v}(Q^2)x^4 \right. 
\\ \nn &+& \left. F_{d_v}(Q^2)x^6\right] \, ,
\\ x \, \bar{d}(x,Q^2) &=& A_{\bar{d}}(Q^2) \, x^{B_{\bar{d}}(Q^2)} \, (1-x)^{C_{\bar{d}}(Q^2)} 
\\ \nn &-& \Theta(x_{C,\bar{d}}-x) A'_{\bar{d}}(Q^2) \, x^{B'_{\bar{d}}(Q^2)} \, (1-x)^{C'_{\bar{d}}(Q^2)}
\label{dBAR}
\\ \nn &\times& \left[1+D'_{\bar{d}}(Q^2)x^2+E'_{\bar{d}}(Q^2)x^4\right] \, ,
\eeqn
with $x_{C,\bar{d}} = 0.04$. We deemed this value to be the best choice to describe the PDFs, by repeating the procedure for a range of $x_{C,\bar{d}}$ and selecting the optimal. Then, the $d$-quark distribution can be recovered by computing
\beq
d(x,Q^2) = d_v(x,Q^2) + \bar{d}(x,Q^2) \, .
\label{d}
\eeq

\begin{figure}[htb]
\centering
\includegraphics[scale=0.5]{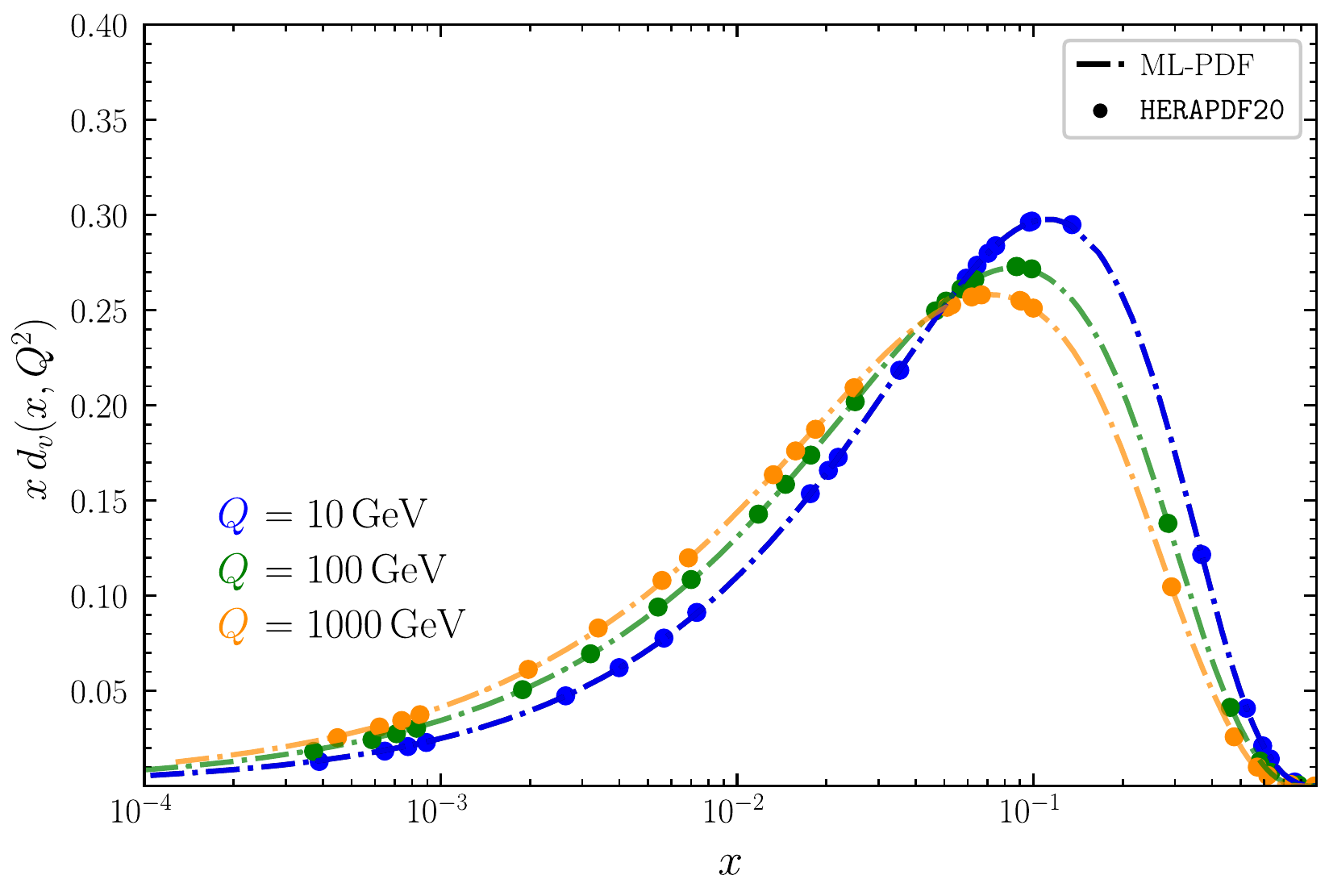}\,
\includegraphics[scale=0.5]{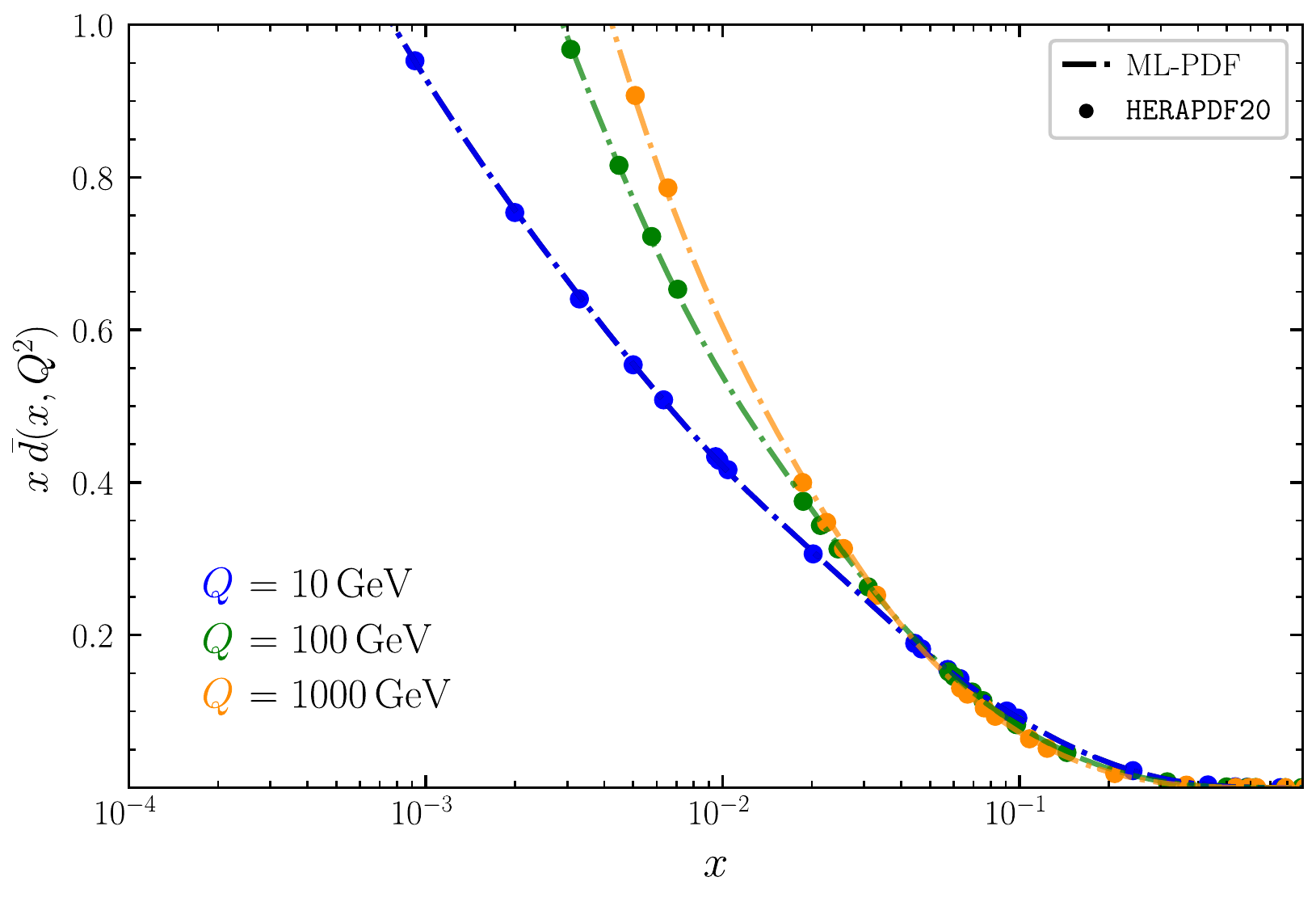}
\caption{Comparison between our analytic ML-PDF approximations (dashed lines) and \texttt{HERAPDF20} (solid dots), for $x \, d_v$ (upper plot) and $x \, \bar{d}$ (lower plot), at three different values of $Q$.}
\label{fig:DOWNdistributions}
\end{figure}

{In Fig. \ref{fig:DOWNdistributions}, we present a comparison between our analytic ML-PDF approximations to $x\, d_v(x,Q^2)$ (upper plot) and $x\, \bar{d}(x,Q^2)$ (lower plot) and the \texttt{HERAPDF20$\_$NLO$\_$EIG} PDF set. We appreciate a very good agreement for different values of $Q$.}

\subsection{Up quark distributions}
\label{ssec:UP}
{The optimal fit was found by considering $u_{v}$ and $\bar{u}$ distributions. To this end, we proposed} 
\beqn
x \, u_v(x,Q^2) &=& A_{u_v}(Q^2) \, x^{B_{u_v}(Q^2)} \, (1-x)^{C_{u_v}(Q^2)} 
\label{uVALENCIA}
\\ \nn &\times& \big[1+D_{u_v}(Q^2)x+E_{u_v}(Q^2)x^2 
\\ \nn &+& F_{u_v}(Q^2)x^3 + G_{u_v}(Q^2)x^4 + H_{u_v}(Q^2)x^5\big] \,,\eeqn
{ to approximate $u_v$, whilst a slightly more complicated ansatz was used for $\bar{u}$. In that case, we split the analysis into two regions:
\beqn
R_1 &=& \{Q_0 \leq Q \leq 10 \, {\rm GeV}\} \, ,
\\ R_2 &=& \{10 \, {\rm GeV} \leq Q \leq 1000 \, {\rm GeV}\} \, .
\eeqn
Then, we defined 
\beqn
\nn x \, \bar{u}(x,Q^2) &=& A_{\bar{u}}(Q^2) \, x^{B_{\bar{u}}(Q^2)} \, (1-x)^{C_{\bar{u}}(Q^2)} 
\\ \nn &\times& \big[1+D_{\bar{u}}(Q^2)x+E_{\bar{u}}(Q^2)x^2 
\\ \nn &+& F_{\bar{u}}(Q^2)x^3 + G_{\bar{u}}(Q^2)x^4 + H_{\bar{u}}(Q^2)x^5\big] \, ,
\label{ubarR1}
\eeqn
for $Q \in R_1$ and 
\beqn
\nn x \, \bar{u}(x,Q^2) &=& A_{\bar{u}}(Q^2) \, x^{B_{\bar{u}}(Q^2)} \, (1-x)^{C_{\bar{u}}(Q^2)} 
\\ \nn &\times& \big[1+D_{\bar{u}}(Q^2)x+E_{\bar{u}}(Q^2)x^3 \big]
\\ \nn &-& \Theta(x_{C,\bar{u}}-x) A'_{\bar{u}}(Q^2) \, x^{B'_{\bar{u}}(Q^2)} 
\\ &\times& (1-x)^{C'_{\bar{u}}(Q^2)}[1+D'_{\bar{u}}(Q^2) x^2] \, ,
\label{ubarR2}
\eeqn
for $Q \in R_2$. Here, we used the cut $x_{C,\bar{d}} = 0.01$, inspired by the gluon PDF parametrization defined by \texttt{HERAPDF}. After fitting these two distributions, we can define 
\beq
u(x,Q^2) = \bar{u}(x,Q^2) + u_v(x,Q^2) \, ,
\label{u}
\eeq
and obtain the $u$-quark distribution.}

\begin{figure}[htb]
\centering
\includegraphics[scale=0.5]{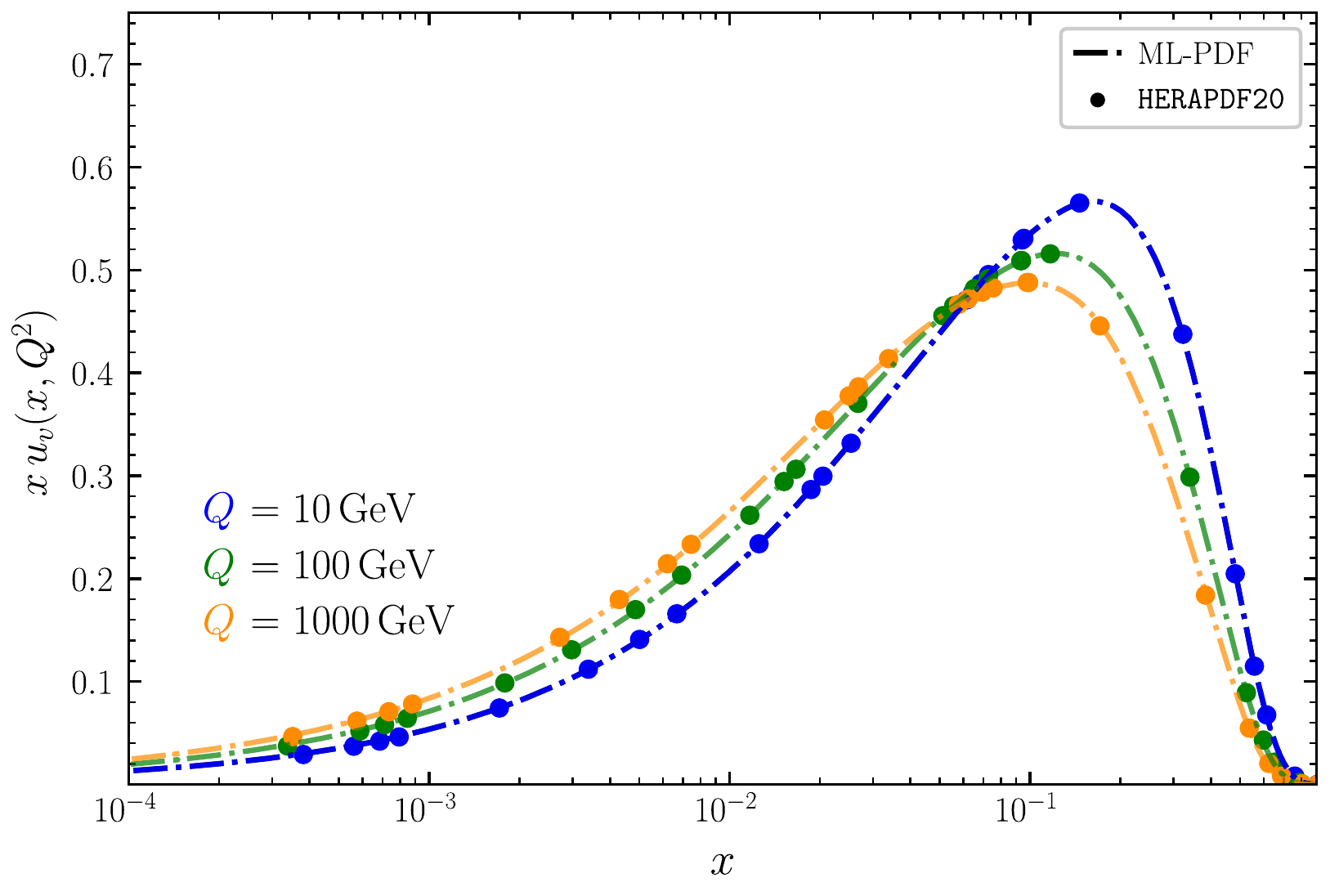} \,
\includegraphics[scale=0.5]{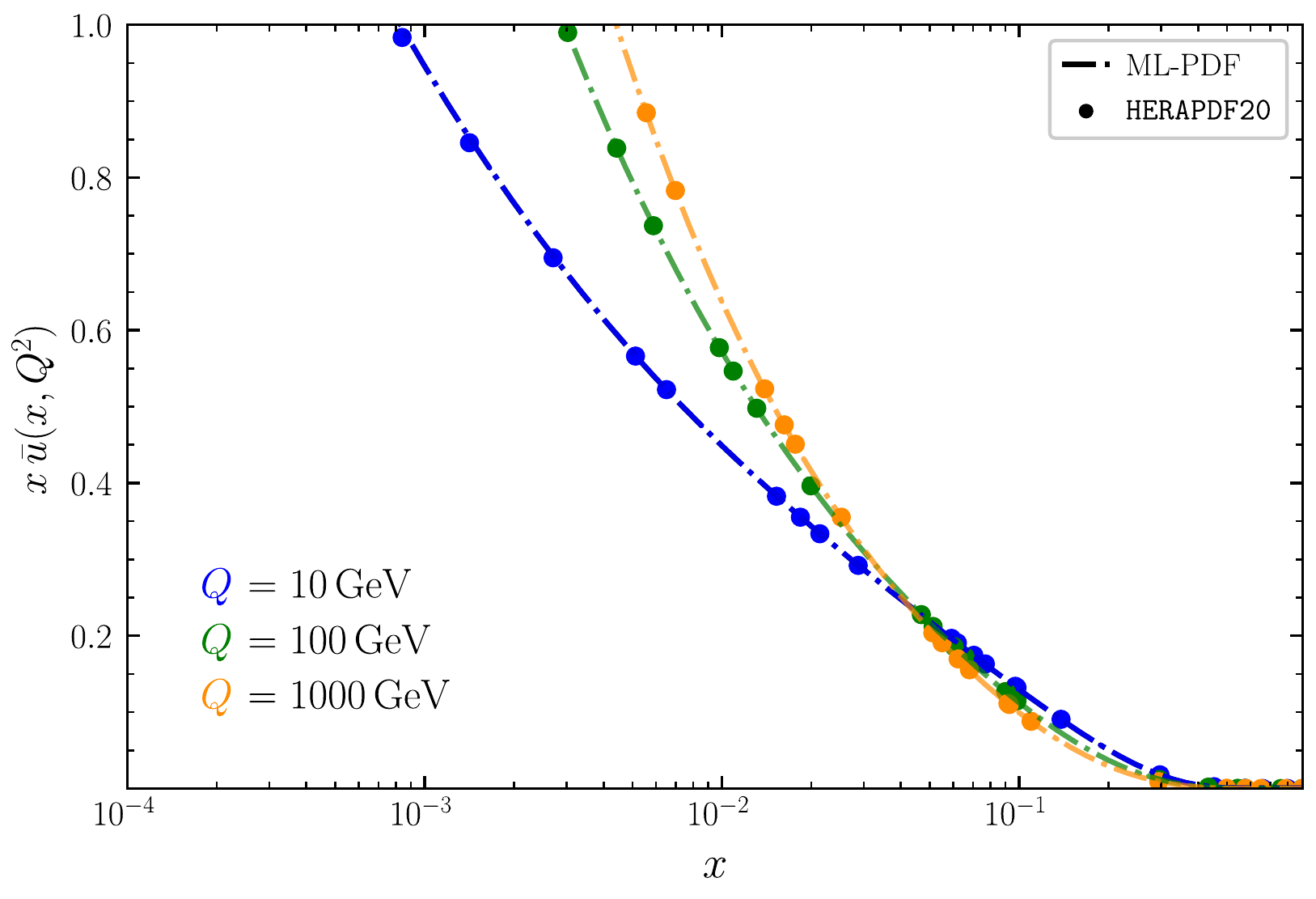}
\caption{Comparison between our analytic ML-PDF approximations (dashed lines) and \texttt{HERAPDF20} (solid dots), for $x \, u_v$ (upper plot) and $x \, \bar{u}$ (lower plot), at three different values of $Q$.}
\label{fig:UPdistributions}
\end{figure}

We show the comparison between our analytic ML-PDF approximations to { $x\, u_v(x,Q^2)$} (upper plot) and { $x\, \bar{u}(x,Q^2)$} (lower plot) w.r.t. \texttt{HERAPDF20$\_$NLO$\_$EIG} PDF set in { Fig. \ref{fig:UPdistributions}}. Again, the agreement is excellent for the complete range of $Q$ considered in this analysis. 

\subsection{Strange quark distribution}
\label{ssec:strange}
For the strange quark we took the usual assumption that $\bar{s}\equiv s$, a common choice for PDFs at NLO that is also imposed for charm and bottom PDFs. After unsuccessfully trying several functional forms based on Euler beta functions, we noticed that the behaviors below and above $Q = 4.5$ GeV were slightly different. Thus, we used this threshold and split the $Q$ analysis into two regions: 
\beqn
R_1 &=& \{Q_0 \leq Q \leq 4.5 \, {\rm GeV}\} \, ,
\\ R_2 &=& \{4.5 \, {\rm GeV} \leq Q \leq 1000 \, {\rm GeV}\} \, .
\eeqn
We defined 
\beqn
\nn x \, s(x,Q^2) &=& A_{s}(Q^2) \, x^{B_{s}(Q^2)} \, (1-x)^{C_{s}(Q^2)} 
\\ \nn &-& \Theta(x_{C,s}(R_1)-x) A'_{s}(Q^2) \, x^{B'_{s}(Q^2)} 
\\ &\times&  (1-x)^{C'_{s}(Q^2)} \, ,
\label{sR1}
\eeqn
and 
\beqn
\nn x \, s(x,Q^2) &=& A_{s}(Q^2) \, x^{B_{s}(Q^2)} \, (1-x)^{C_{s}(Q^2)} 
\\ \nn &-& \Theta(x_{C,s}(R_2)-x) A'_{s}(Q^2) \, x^{B'_{s}(Q^2)} 
\\ &\times& (1-x)^{C'_{s}(Q^2)}[1+D'_s(Q^2) x^2] \, ,
\label{sR2}
\eeqn
where $x_{C,s}(R_1)=0.25$ and $x_{C,s}(R_2)=0.1$ in $R_1$ and $R_2$, respectively. Once again, these values were determined by exploring a range of values and choosing the bests ones (i.e., the ones that gave a smaller value of the integral error).

\begin{figure}[htb]
\centering
\includegraphics[scale=0.5]{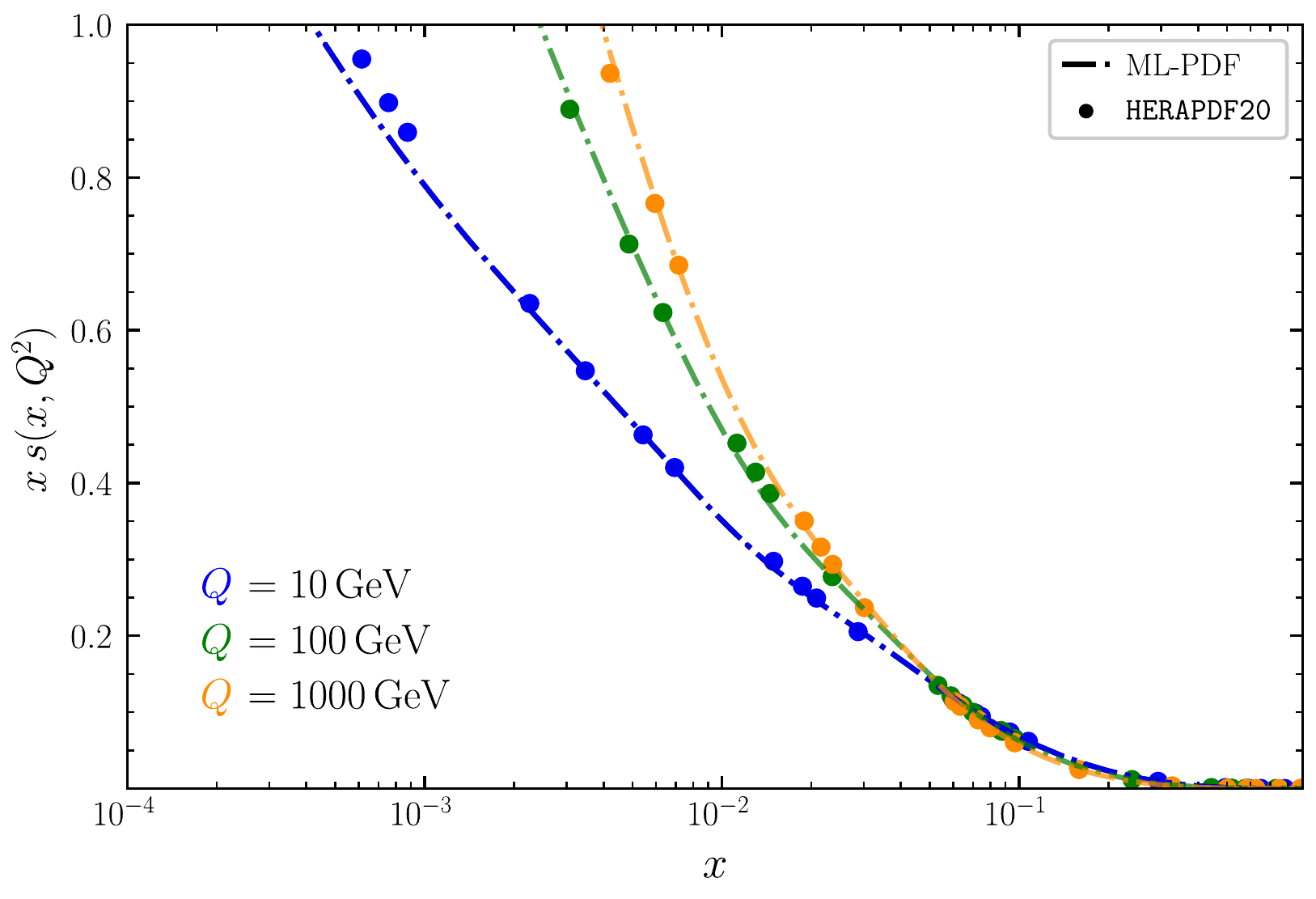}
\caption{Comparison between our analytic ML-PDF strange distribution approximation (dashed lines) and \texttt{HERAPDF20} (solid dots). We considered three different values of $Q$.}
\label{fig:Sdistribution}
\end{figure}

\begin{figure}[htb]
\centering
\includegraphics[scale=0.5]{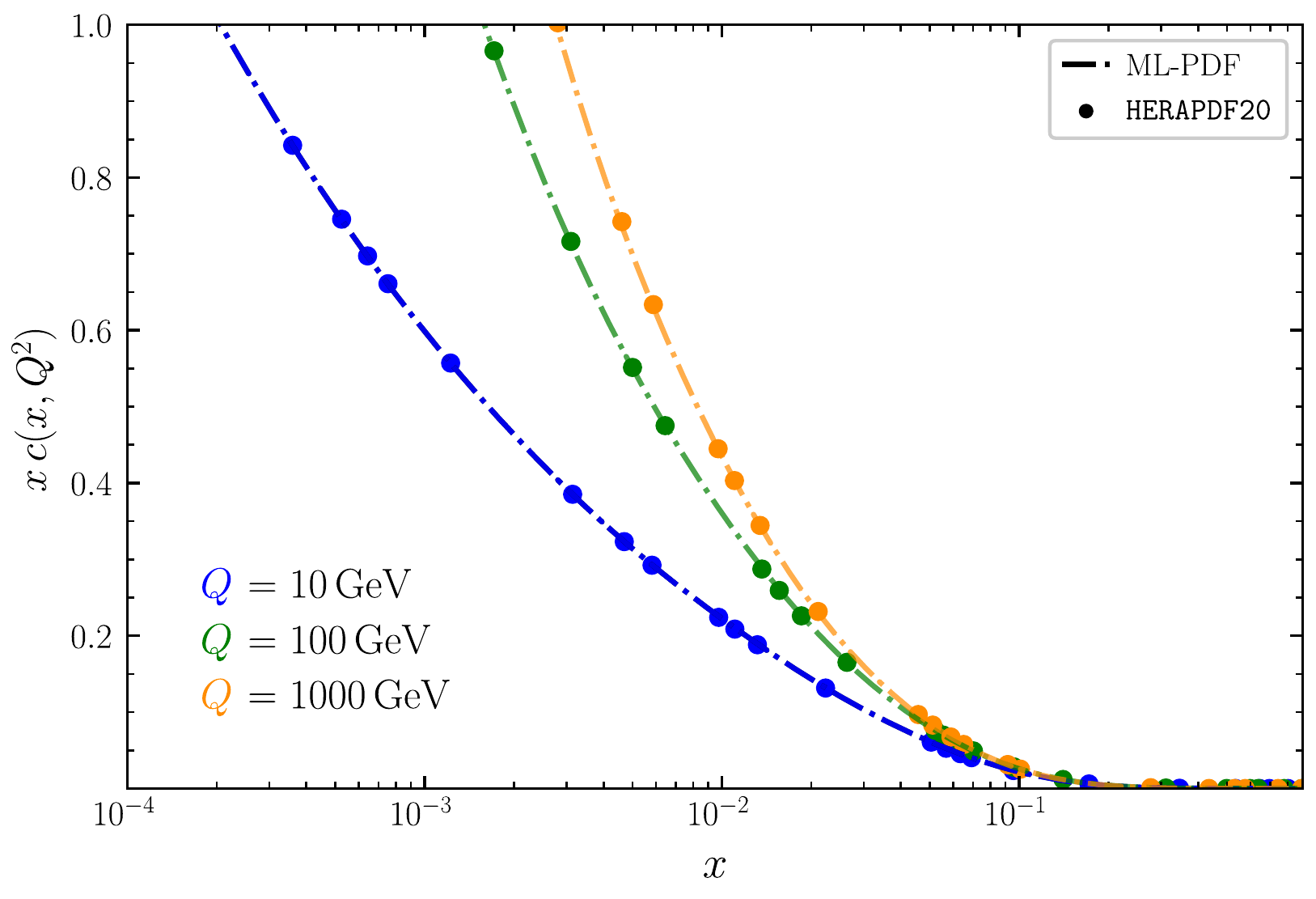}
\caption{Comparison between our analytic ML-PDF charm distribution approximation (dashed lines) and \texttt{HERAPDF20} (solid dots). We considered three different values of $Q$.}
\label{fig:Cdistribution}
\end{figure}

In Fig. \ref{fig:Sdistribution}, we show our analytic ML-PDF approximation to $x\, s(x,Q^2)$ (dashed lines) w.r.t. the corresponding strange PDF from \texttt{HERAPDF20$\_$NLO$\_$EIG} (solid dots). At low-$x$ (i,.e. below $10^{-3}$), our fit slightly undershoots the prediction from \texttt{HERAPDF}. Still, the agreement is very good for $x > 10^{-3}$, and, in particular, for different values of $Q$. 

\subsection{Charm quark distribution}
\label{ssec:Charm}
As mentioned above, we considered the heavy quarks as generated radiatively by the gluons, and thus fixed their distributions to zero when the scale is below their respective production threshold. Above it, in the case of the charm quark, we split the analysis in two different regions in $Q$. Explicitly, we considered: 
\beqn
R_1 &=& \{1.47 \, {\rm GeV} < Q \leq 3 \, {\rm GeV}\} \, ,
\\ R_2 &=& \{3 \, {\rm GeV} \leq Q \leq 1000 \, {\rm GeV}\} \, ,
\eeqn
since $m_c^{\rm pole}=1.47$ GeV according to \texttt{HERAPDF2.0} fit including NLO QCD corrections \cite{Zhang:2015tuh}. Thus, we defined 
\beqn
\nn x \, c(x,Q^2) &=& A_{c}(Q^2) \, x^{B_{c}(Q^2)} \, (1-x)^{C_{c}(Q^2)} 
\\ \nn &\times& [1+D_{c}(Q^2) x^2]
\\ \nn &-& \Theta(x_{C,c}(R_1)-x) A'_{c}(Q^2) \, x^{B'_{c}(Q^2)} 
\\ &\times&  (1-x)^{C'_{c}(Q^2)} \, [1+D'_{c}(Q^2) x^2] \, ,
\label{cR1}
\eeqn
and 
\beqn
\nn x \, c(x,Q^2) &=& A_{c}(Q^2) \, x^{B_{c}(Q^2)} \, (1-x)^{C_{c}(Q^2)} 
\\ \nn &\times& (1+D_{c}(Q^2) x+E_c(Q^2)x^2)
\\ \nn &-& \Theta(x_{C,c}(R_2)-x) A'_{c}(Q^2) \, x^{B'_{c}(Q^2)} 
\\ &\times& (1-x)^{C'_{c}(Q^2)}(1+D'_c(Q^2) x^2) \, ,
\label{cR2}
\eeqn
where $x_{C,c}(R_1)=0.1$ and $x_{C,c}(R_2)=0.05$ in $R_1$ and $R_2$, respectively.

We present a comparison between our analytic ML-PDF approximation to $x\, c(x,Q^2)$ (dashed lines) versus the values provided by \texttt{HERAPDF20$\_$NLO$\_$EIG} (solid dots) in Fig. \ref{fig:Cdistribution}. The agreement is impressive, both in $x$ and $Q$. 

\begin{figure}[htb]
\centering
\includegraphics[scale=0.5]{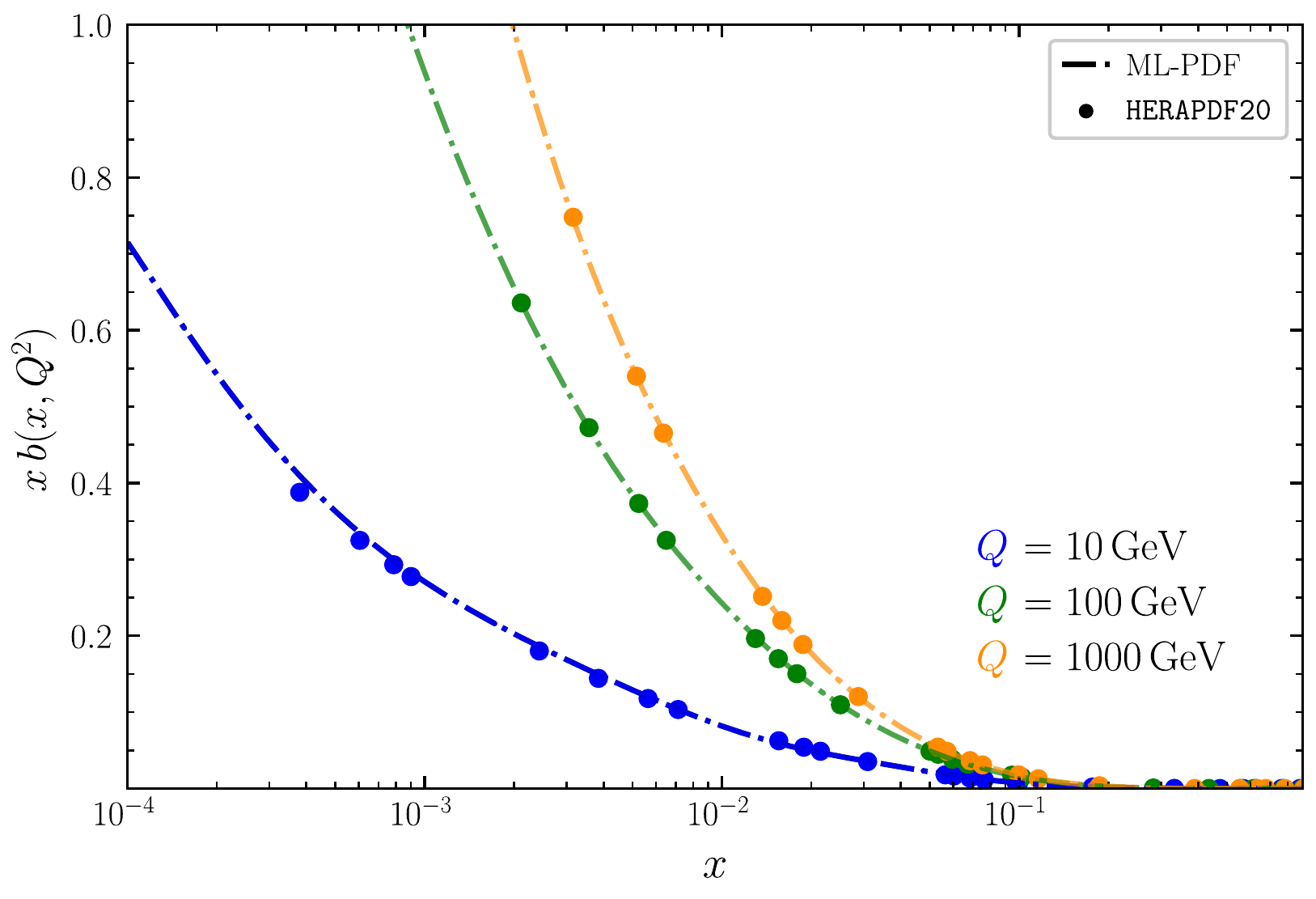}
\caption{Comparison between our analytic bottom ML-PDF approximation (dashed lines) and \texttt{HERAPDF20} (solid dots). We considered three different values of $Q$.}
\label{fig:Bdistribution}
\end{figure}

\subsection{Bottom quark distribution}
\label{ssec:Bottom}
Finally, we studied the bottom quark distribution. Similarly to the case of the charm quark, we started the fitting for values of $Q$ above the mass of the bottom, i.e., $Q > 4.5$ GeV. We noticed that the quality of our fit significantly increased if we divided the $Q$ range into two regions: 
\beqn
R_1 &=& \{4.5 \, {\rm GeV} \leq Q \leq 15 \, {\rm GeV}\} \, ,
\\ R_2 &=& \{15 \, {\rm GeV} \leq Q \leq 1000 \, {\rm GeV}\} \, .
\eeqn
As we did for the other flavours, we then proposed
\beqn
\nn x \, b(x,Q^2) &=& A_{b}(Q^2) \, x^{B_{b}(Q^2)} \, (1-x)^{C_{b}(Q^2)} 
\\ \nn &\times& [1+D_{b}(Q^2) x^2]
\\ \nn &-& \Theta(x_{C,b}(R_1)-x) A'_{b}(Q^2) \, x^{B'_{b}(Q^2)} 
\\ &\times&  (1-x)^{C'_{b}(Q^2)} \, [1+D'_{b}(Q^2) x^2] \, ,
\label{bR1}
\eeqn
and 
\beqn
\nn x \, b(x,Q^2) &=& A_{b}(Q^2) \, x^{B_{b}(Q^2)} \, (1-x)^{C_{b}(Q^2)} 
\\ \nn &\times& [1+D_{b}(Q^2) x+E_b(Q^2)x^2]
\\ \nn &-& \Theta(x_{C,b}(R_2)-x) A'_{b}(Q^2) \, x^{B'_{b}(Q^2)} 
\\ &\times& (1-x)^{C'_{b}(Q^2)}[1+D'_b(Q^2) x^2] \, ,
\label{bR2}
\eeqn
where $x_{C,b}(R_1)=0.1$ and $x_{C,b}(R_2)=0.05$ in $R_1$ and $R_2$, respectively.

As for the other flavours, in Fig. \ref{fig:Bdistribution}, we show our analytic ML-PDF approximation to $x\, b(x,Q^2)$ (dashed lines) w.r.t. the corresponding bottom PDF from \texttt{HERAPDF20$\_$NLO$\_$EIG} (solid dots). The agreement is very good for different values of $Q$, from $10$ to $500$ GeV. Even if some discrepancies arise for $x < 5 \cdot 10^{-4}$, we can state that our formulae globally provides a reliable approximation to the bottom PDF.


\section{Efficiency and quality benchmarks}
\label{sec:Comparison}
In this section we discuss quantitatively the quality of the ML-PDF approximations found in Sec. \ref{sec:Results} and compare the time required to compute some chosen observables. 

\begin{figure}[htb]
\centering
\includegraphics[scale=0.5]{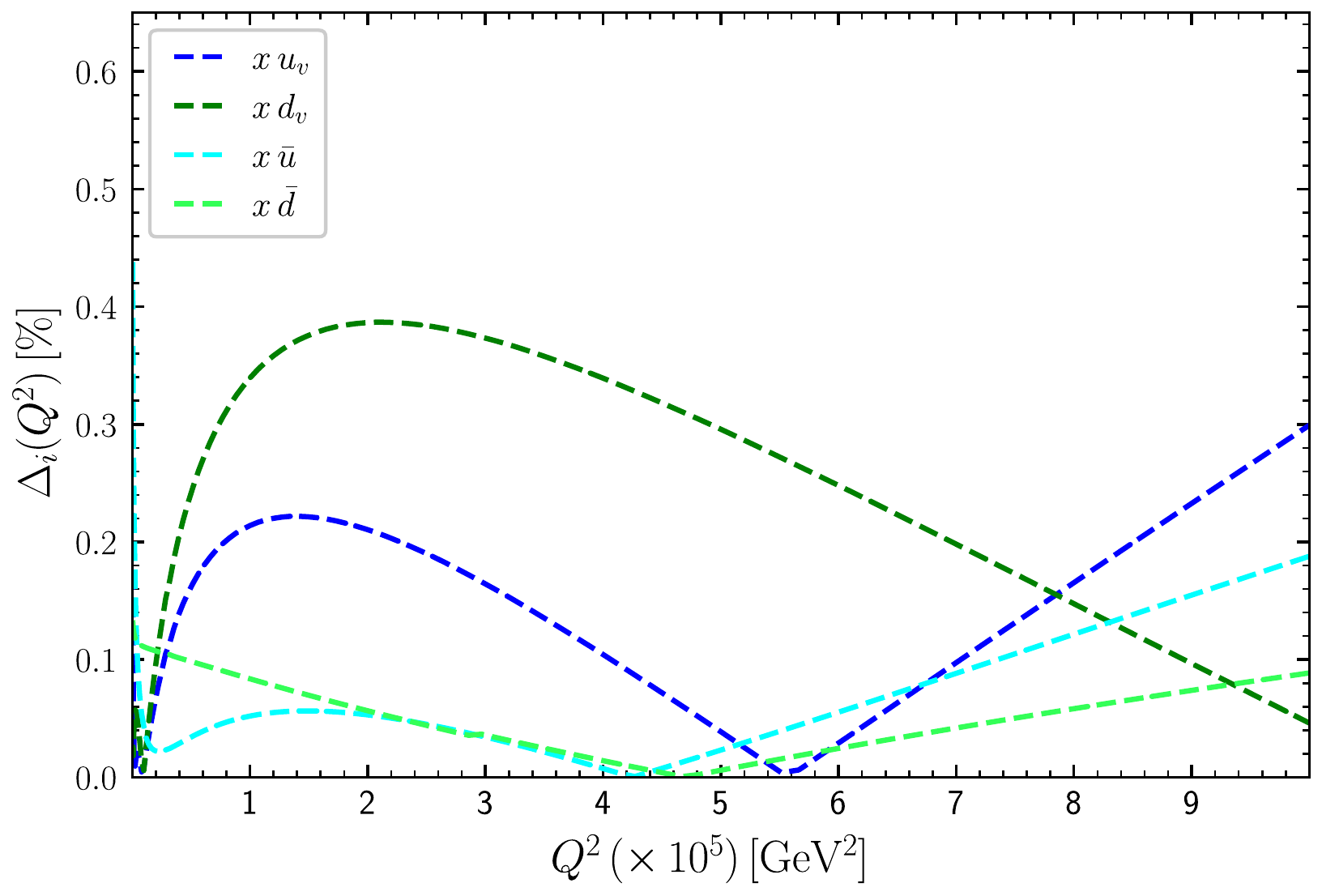} \,
\includegraphics[scale=0.5]{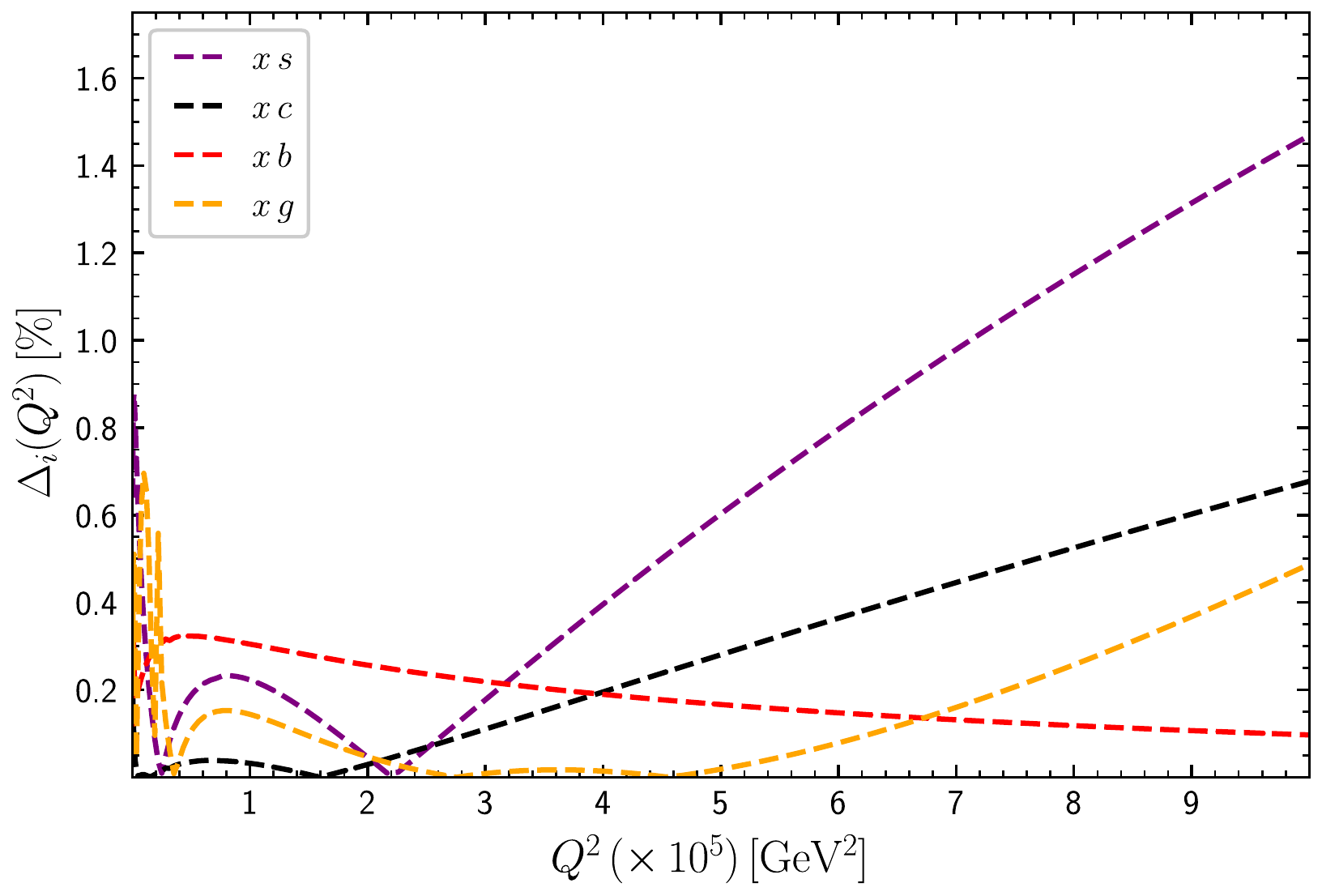}
\caption{Integral error for our analytic ML-PDF approximations w.r.t. \texttt{HERAPDF} distributions. In the upper plot, we show up and down quark distributions, whilst $s$, $c$, $b$ and gluon PDF are presented in the lower plot.}
\label{fig:ErroresIntegrales}
\end{figure}

In order to estimate the discrepancies between our analytic ML-PDFs and the original \texttt{HERAPDF2.0} distributions in a phenomenologically-relevant way, we rely on our definition of integral error given in Eq. (\ref{eq:IntegralError}). In Fig. \ref{fig:ErroresIntegrales}, we show that the integral error for almost all the distributions (i.e. $u_v$, $d_v$, $\bar{u}$, $\bar{d}$, $c$, $b$ and $g$) is below $0.5 \, \%$ in the $Q \in [Q_0,1000 \, {\rm GeV}]$ range. For the strange quark PDF, we appreciate a larger deviation at higher values of $Q$, reaching up to $1.5 \, \%$ error for $Q \approx 1000$ GeV. { Although not present in the plots, we want to highlight that the integral error for $u$ and $d$ is also well under control, being below percent level for the whole range of $Q$-values explored. This is expected from Eqs. (\ref{d}) and (\ref{u}), propagating the errors shown in Fig. \ref{fig:ErroresIntegrales}.}

\begin{figure}[htb]
\centering
\includegraphics[scale=0.5]{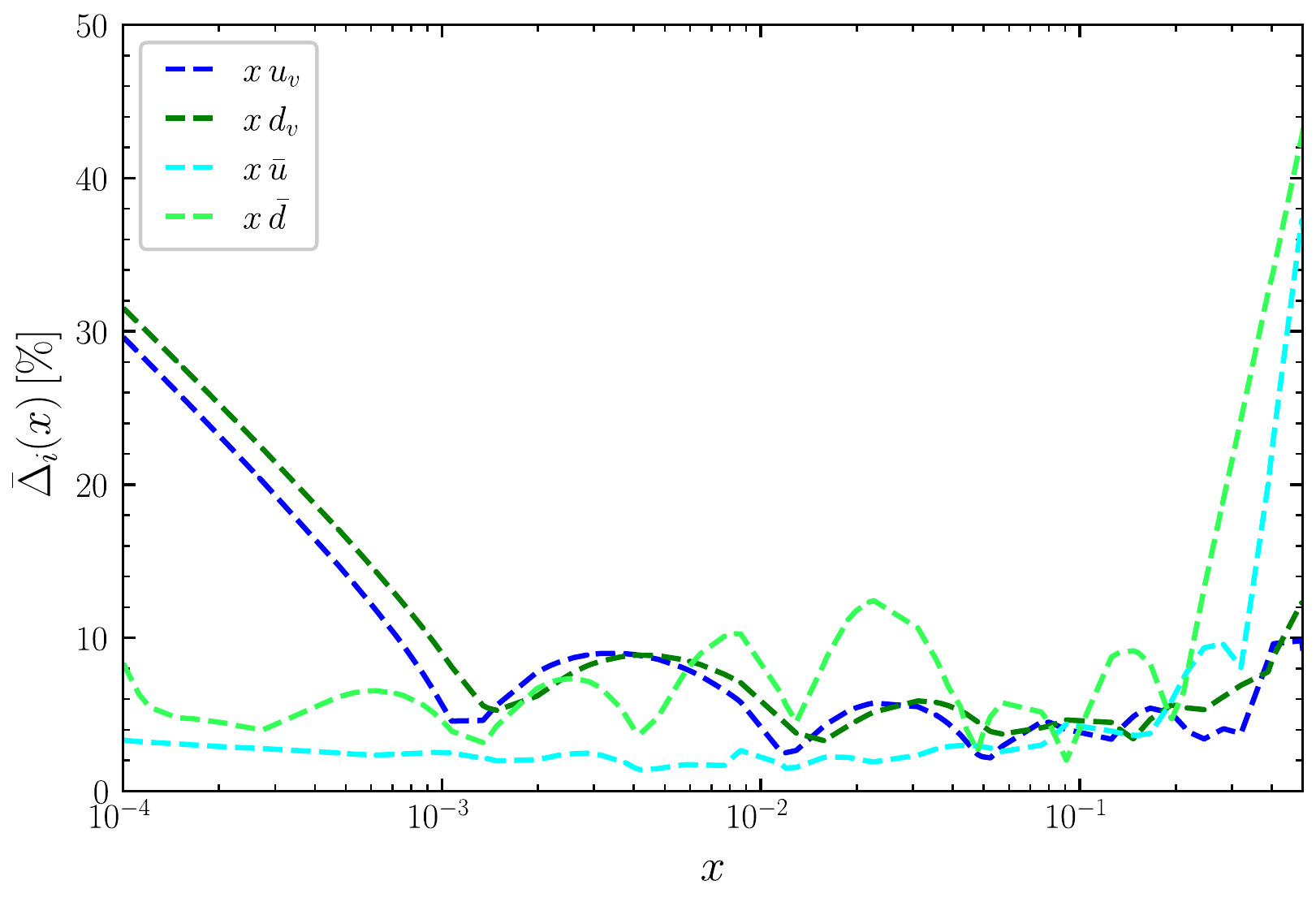} \,
\includegraphics[scale=0.5]{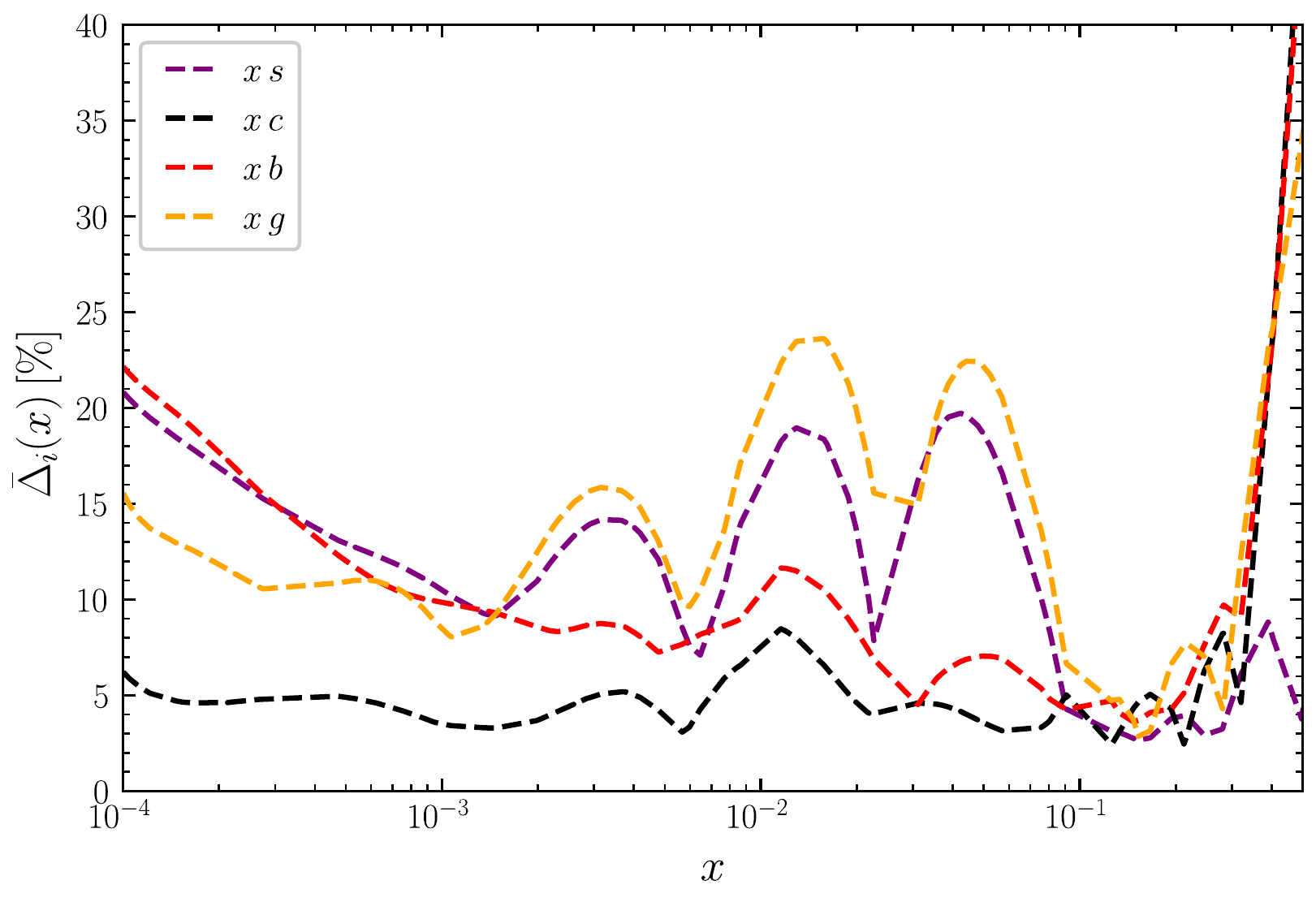}
\caption{ Estimation of the error as a function of $x$, averaged over $Q\in[10\,{\rm GeV},1000\,{\rm GeV}]$ for our analytic ML-PDF approximations w.r.t. \texttt{HERAPDF2.0} distributions. In the upper plot, we show up and down quark distributions, whilst $s$, $c$, $b$ and gluon PDFs are presented in the lower plot.}
\label{fig:ErroresenX}
\end{figure}

{As we explained in Sec. \ref{sec:Methodology}, the integral error is expected to provide a more reliable estimation on the impact of using our ML-PDFs instead of the original PDF set in a physical calculation. Still, for the sake of completeness, we present here an analysis of the approximation error as a function of $x$. For this purpose, we sample with an exponential distribution the range $Q \in [10\, {\rm GeV},1000 \,{\rm GeV}]$ using $N=50,000$ random points. Then, we define
\beq
\bar{\Delta}_i (x) = \frac{1}{N} \sum_{j=1}^{N} \, \Bigg\vert 1-\frac{f_i^{\rm ML}(x,Q_j^2)}{f_i^{\rm HERA}(x,Q_j^2)} \Bigg\vert \, ,
\eeq
as an estimator of the error in $x$. This definition is similar to the \emph{error shape} mentioned in Sec. \ref{sec:Methodology}, and corresponds to an average of the relative errors as a function of $x$. In Fig. \ref{fig:ErroresenX}, we show the results for up, down (upper plot), strange, charm, bottom quarks and the gluon (lower plot). For $u_v$, $d_v$, $\bar{u}$ and $\bar{d}$-quarks, $\bar{\Delta}$ is below $10\,\%$ for $x\in[10^{-3},0.3]$. Similarly, for the other quark flavours, the error is below $20\,\%$ in $x\in[10^{-3},0.3]$. For the gluon, the error is slightly larger, ${\cal O}(10-25\,\%)$ in the central $x$ region. All the distributions tend to increase their average relative error for $x>0.3-0.4$, since the PDFs decrease a couple of orders of magnitude for larger values of $x$. Also, the valence distributions show this behavior in the low-$x$ region. In both cases, small absolute discrepancies translate into large relative fluctuations. This effect is shadowed in the definition of integral error, because the contribution of the PDFs in those regions are rather small. 

Another important observation is the presence of oscillations in the relative error in the region $x\approx 0.05-0.1$, particularly for distributions that made use of a \emph{gluon-like} functional form. This behaviour is due to the matching of two Euler beta functions around $x_{C}={\cal O}(0.1)$, which originates a fluctuation since they have different signs. In fact, this justifies the reduced integral error because these fluctuations around the original PDF cancel at integrand level.}

Regarding the time needed for the use of the ML-PDFs and \texttt{LHAPDF}, we found a significant difference between simply calling the distributions and actually implementing them in a cross-section calculation. The results presented in the rest of this section were obtained using a desktop PC with a 16-core Intel i7-13700 processor. It is important to remark that nothing has been parallelised. 

\subsection{Run time}
\label{ssec:runtime}

We start by presenting, in Tab. (\ref{tab:speed}), the run-time difference when performing a call to our PDFs and \texttt{LHAPDF}, for sets of points in the $(x,Q^2)$ space. As we can appreciate, for $N_{points}\leq 10^{3-4}$ the gain is quite substantial. This is due to the fact that calling \texttt{LHAPDF} has an overhead time for loading and reading the grid, which albeit small, dominates the total run time when one evaluates a small number of points. Therefore, if we were interested in evaluating a handful of points we would be wise in choosing our ML-PDFs over \texttt{LHAPDF}. As we move to larger number of points $N_{points}\approx 10^5$ the weight of this overhead starts to dilute, and once we pass the mark $N_{points}\approx 10^{6}$, we reach a region where we are essentially comparing only the execution time of interpolation versus evaluation. 
{ In the last column we display the percentual time gain, defined as
\begin{equation}
\text{gain(\%) }=100\times \frac{\text{time}_{\text{LHAPDF}}-\text{time}_{\text{ML-PDF}}}{\text{time}_{\text{LHAPDF}}} \, .
\end{equation}
A negative gain would mean that the interpolation is faster than the direct evaluation. As can be seen from the table, the gain seems to approach a plateau around $50\%$ (i.e. our ML-PDF are two times faster than \texttt{LHAPDF}), which is quite sizable. 

\begin{table}[h!]
\begin{center}
\begin{tabular}{|c |c |c |c|} 
 \hline
 \, $N_{points}$ \, & \, \texttt{LHAPDF} (s) \, & \, ML-PDFs (s)\,  & \, Gain($\%$) \, \\ [0.5ex] 
 \hline\hline
 $10^3$ & $3.76 \cdot 10^{-2}$ & $2.92 \cdot 10^{-4}$ & 99.22 \\ 
 \hline
 $10^4$ & $4.20 \cdot 10^{-2}$ & $2.50 \cdot 10^{-3}$ & 94.05 \\ 
 \hline
 $10^5$ & $8.94 \cdot 10^{-2}$ & $2.50 \cdot 10^{-2}$ & 72.10 \\ 
 \hline
 $10^6$ & $0.56$ & $0.25$ & 55.46 \\ 
 \hline
 $10^7$ & $5.25$ & $2.50$ & 52.49 \\ 
  \hline
  $10^8$ & $52.04$ & $24.92$ & 52.11 \\ 
 \hline
\end{tabular}
\caption{Comparison of the time (in seconds) required to compute $N_{points}$ evaluations of \texttt{HERAPDF2.0} within \texttt{LHAPDF} framework, and our ML-PDF analytic approximation.}
\label{tab:speed}
\end{center}
\end{table}

From these numbers it appears that, for a code requiring Monte-Carlo integration, it would be beneficial to use something akin to our ML-PDFs. We further investigate this in the next sub-section.}

{
\subsection{Validity of the sum rules}
\label{ssec:SumRules}
One important physically-motivated cross-check of our analytic ML-PDFs are the sum rules. For any energy scale $Q$, the sum rules are given by
\beqn
S_1(Q^2)=\int_0^1 \, dx \, u_v(x,Q^2) &=& 2 \, , 
\label{eq:UVSum}
\\ S_2(Q^2)=\int_0^1 \, dx \, d_v(x,Q^2) &=& 1 \, ,
\label{eq:DVSum}
\eeqn
for the up and down valence quarks, and
\beq
\int_0^1 \, dx \, [f_i(x,Q^2)-\bar{f}_i(x,Q^2)] = 0 \, ,
\label{eq:SeaSum}
\eeq
for all other quarks. Also, the PDFs must fulfill that the total momenta carried by the constituents partons equal the momentum of the hadron. This is encoded in the momentum sum rule
\beq
S_3(Q^2)=\int_0^1 \, dx \, x \left[ g(x,Q^2) + \sum_{i\in\{q,\bar{q}\}} f_i(x,Q^2) \, \right] = 1 \, ,
\label{eq:MomentumSR}
\eeq
where the sum is carried out over all the active flavours of quarks. These constraints are often imposed when performing the PDF extraction from fits to experimental data. In the case of \texttt{HERAPDF2.0}, Eq. (\ref{eq:SeaSum}) is automatically fulfilled by construction (no $q-\bar{q}$ distinction for sea quarks), and this also holds for our analytic ML-PDFs.

\begin{figure}[htb]
\centering
\includegraphics[scale=0.45
]{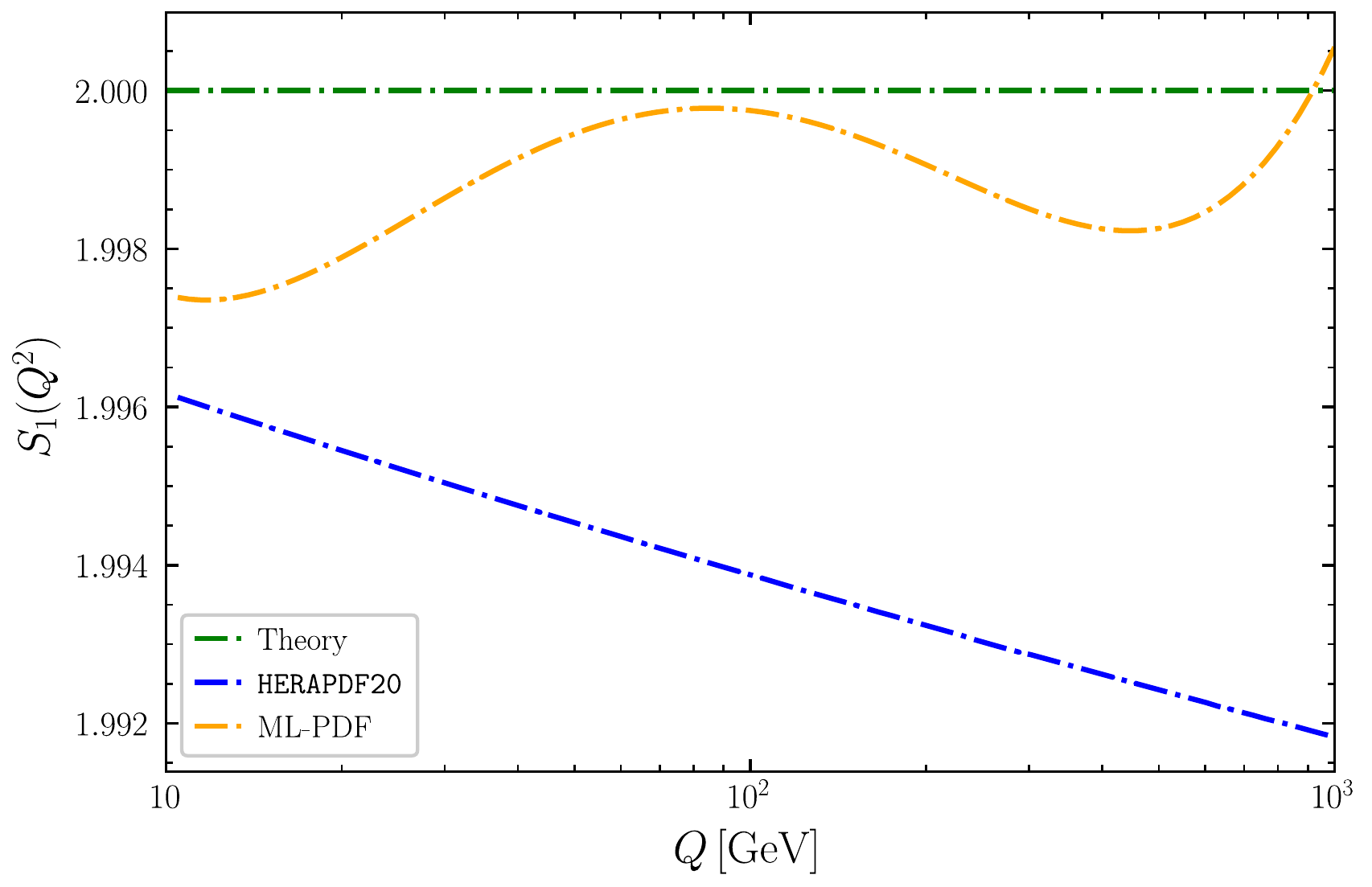} \,
\includegraphics[scale=0.45 ]{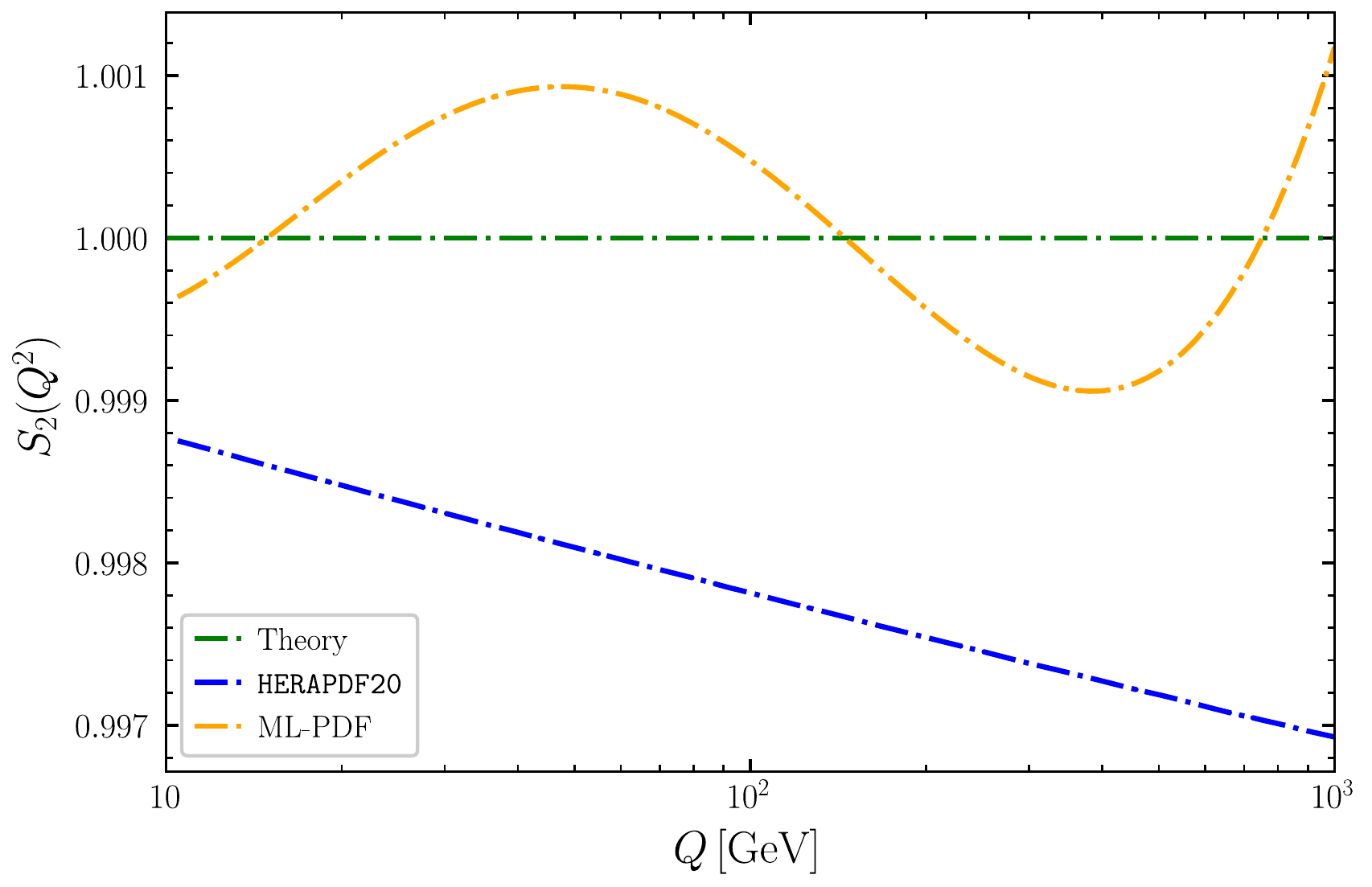}
\caption{Sum rules for up (upper) and down (lower) valence distributions as a function of $Q$. We show the values calculated with \texttt{HERAPDF2.0} (blue) and our analytic ML-PDF (orange), and also the theoretical exact value (green).}
\label{fig:SumRule12}
\end{figure}

\begin{figure}[htb]
\centering
\includegraphics[scale=0.45]{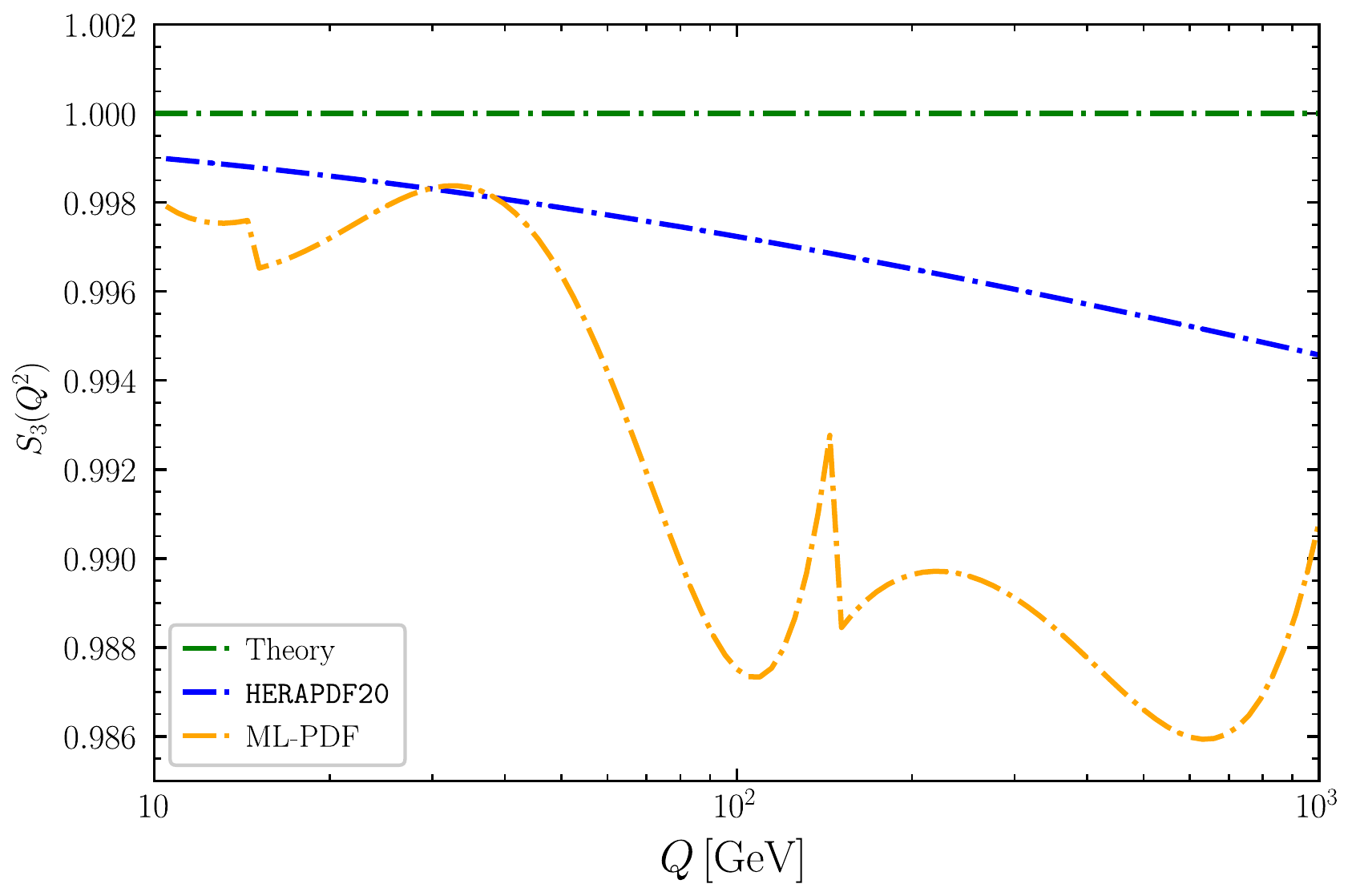}
\caption{Total momentum conservation rule, as a function of $Q$. We show the values calculated with \texttt{HERAPDF2.0} (blue) and our analytic ML-PDF (orange), and also the theoretical exact value (green).}
\label{fig:SumRule3}
\end{figure}

In Fig. \ref{fig:SumRule12} we compare the result of Eqs. (\ref{eq:UVSum}) and (\ref{eq:DVSum}) for \texttt{HERAPDF2.0} (blue) and our analytic ML-PDFs (orange) w.r.t. the theoretical value (green line). In both cases, we consider $Q \in [10 \, {\rm GeV},1000 \, {\rm GeV}]$ and we find that our analytic ML-PDF leads to results closer to the theoretical value. In concrete, for up valence quark, we get deviations of ${\cal O}(0.1 \, \%)$ for ML-PDFs and ${\cal O}(0.2 \, \%)$ for
\texttt{HERAPDF2.0}, showing the last one a trend to depart from the theoretical value in the high $Q$ region. Similarly, for down valence distributions, the predictions obtained with ML-PDFs oscillate around 1 within a band of ${\cal O}(0.2 \, \%)$, while \texttt{HERAPDF2.0} deviates more than ${\cal O}(0.3 \, \%)$ for higher $Q$-values.

Finally, in Fig. \ref{fig:SumRule3}, we show the results of computing Eq. (\ref{eq:MomentumSR}) with our analytic ML-PDFs (orange) and \texttt{HERAPDF2.0} (blue). In this case, \texttt{HERAPDF2.0} approaches better to the theoretical value, with discrepancies smaller than ${\cal O}(0.5\, \%)$. The predictions obtained with our analytic ML-PDFs shows deviations of ${\cal O}(0.6 \, \%)$ below 60 GeV, reaching up to ${\cal O}(1.5 \, \%)$ for $Q \approx 500$ GeV. This behaviour is driven by gluon PDF, since this parton carries more than $45 \, \%$ of proton momentum and our fit show an increasing error for large $Q$ (see Figs. \ref{fig:ErroresIntegrales} and \ref{fig:ErroresenX}). In any case, as we emphasized in the Introduction, this constitutes a first proof-of-concept and we can claim that our ML-PDFs successfully passed all the cross-check with a percent (and even sub-percent) level precision.}


\subsection{Impact in physical observables}
\label{ssec:PhysicalObservables}
Here, we discuss the impact of using our analytic approximations to the PDFs in two realistic cross-section calculations.

\begin{figure}[htb]
\centering
\includegraphics[scale=0.62]{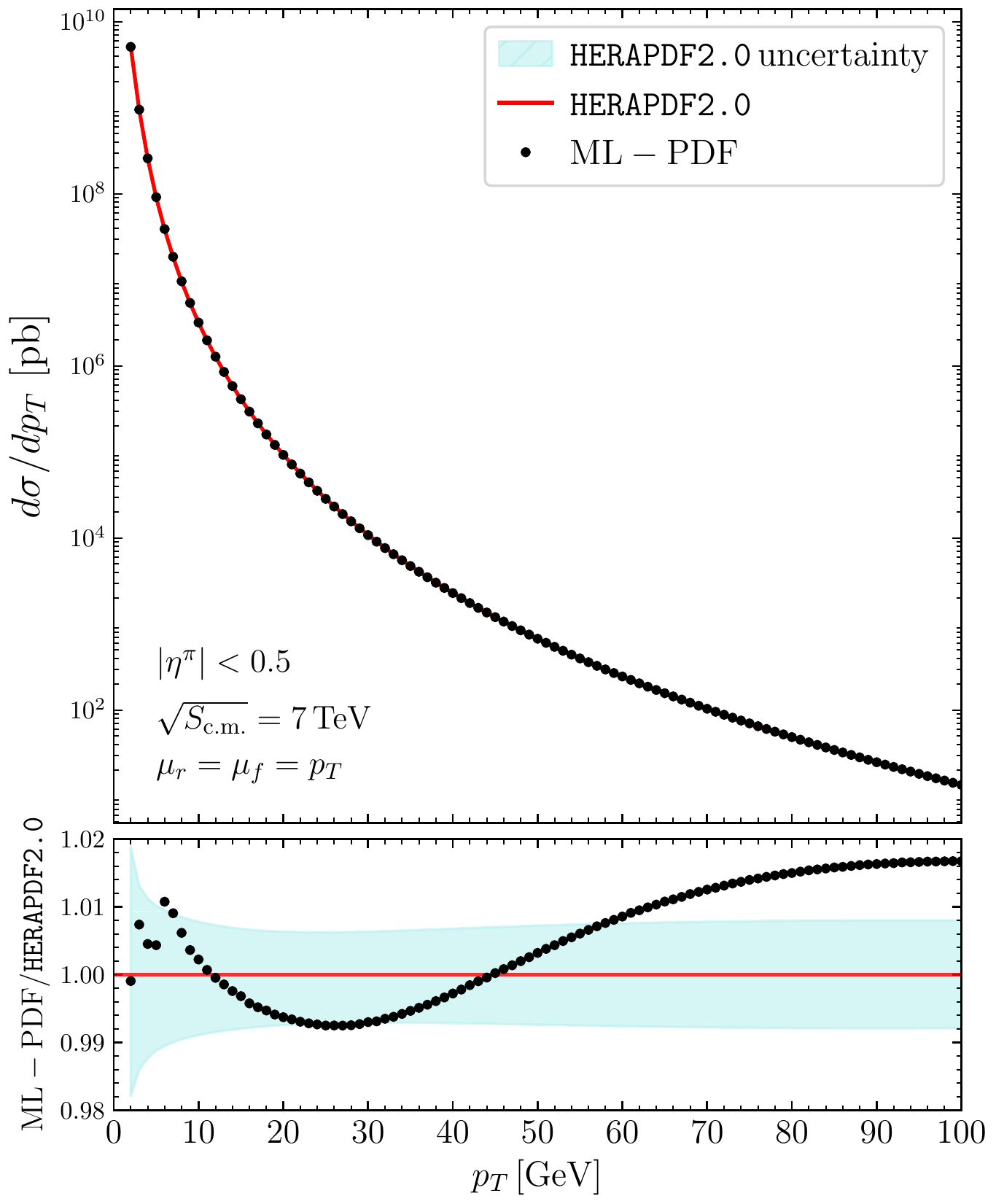}
\caption{Upper plot: differential cross-section for the process $p+p \to \pi$ as a function of $p_T$, including up to NLO QCD corrections. We run the simulation using our analytic ML-PDFs (dots) and \texttt{LHAPDF} with \texttt{HERAPDF2.0} (solid). Lower plot: ratio between the two cross-sections. The discrepancies in the results are well below $2 \, \%$, which is even less than the error introduced by scale variations (not included in this plots).}
\label{fig:final_sih}
\end{figure}

We start by considering pion production in unpolarised proton-proton collisions at $\sqrt{S_{\rm c.m.}}=7$ TeV in the central rapidity region ($|\eta^\pi|<0.5$). These simulations make use of a sequential code based on Ref. \cite{Jager:2002xm}, modified to use the needed PDFs and fragmentation functions (FFs). The Monte-Carlo integrator is \texttt{VEGAS}, with $10^5$ points, the FF set used is \texttt{DEHSS2014} \cite{Hernandez-Pinto:2016cnc}, and all factorization and renormalization scales have been taken to be equal to the transverse momentum of the pion (denoted $p_T$). In the upper plot of Fig. \ref{fig:final_sih}, we present the $p_T$ spectrum using our ML-PDFs (black dots) and \texttt{HERAPDF2.0} (red line), while the lower plot depicts the ratio between these two quantities. The (light-blue) uncertainty band in the lower plot comes from the Hessian set of theoretical uncertainties of \texttt{HERAPDF2.0}; it is also present (but hardly noticeable) in the upper plot. From the lower panel, we can appreciate that the difference between both simulations is well below $2 \, \%$, and mostly stays within the PDF uncertainty bands. This is much smaller than the typical perturbative error of NLO QCD calculations (generally larger than ${\cal O}(20 \, \%)$). Regarding the CPU time, { it took 10 minutes 28 seconds using \texttt{LHAPDF} and 9 minutes 19 seconds with our ML-PDFs}, running on the same system described before. From Tab. \ref{tab:speed}, and considering the number of iterations used, one would expect a much larger gain. The outcome of this run serves to emphasize that the matter is far more subtle. It is worth highlighting that this reduction in the computational time is only due to our optimized ML-PDF, since all the other ingredients of the calculation remain the same in both simulations.

\begin{figure}[htb]
\centering
\includegraphics[scale=0.61]{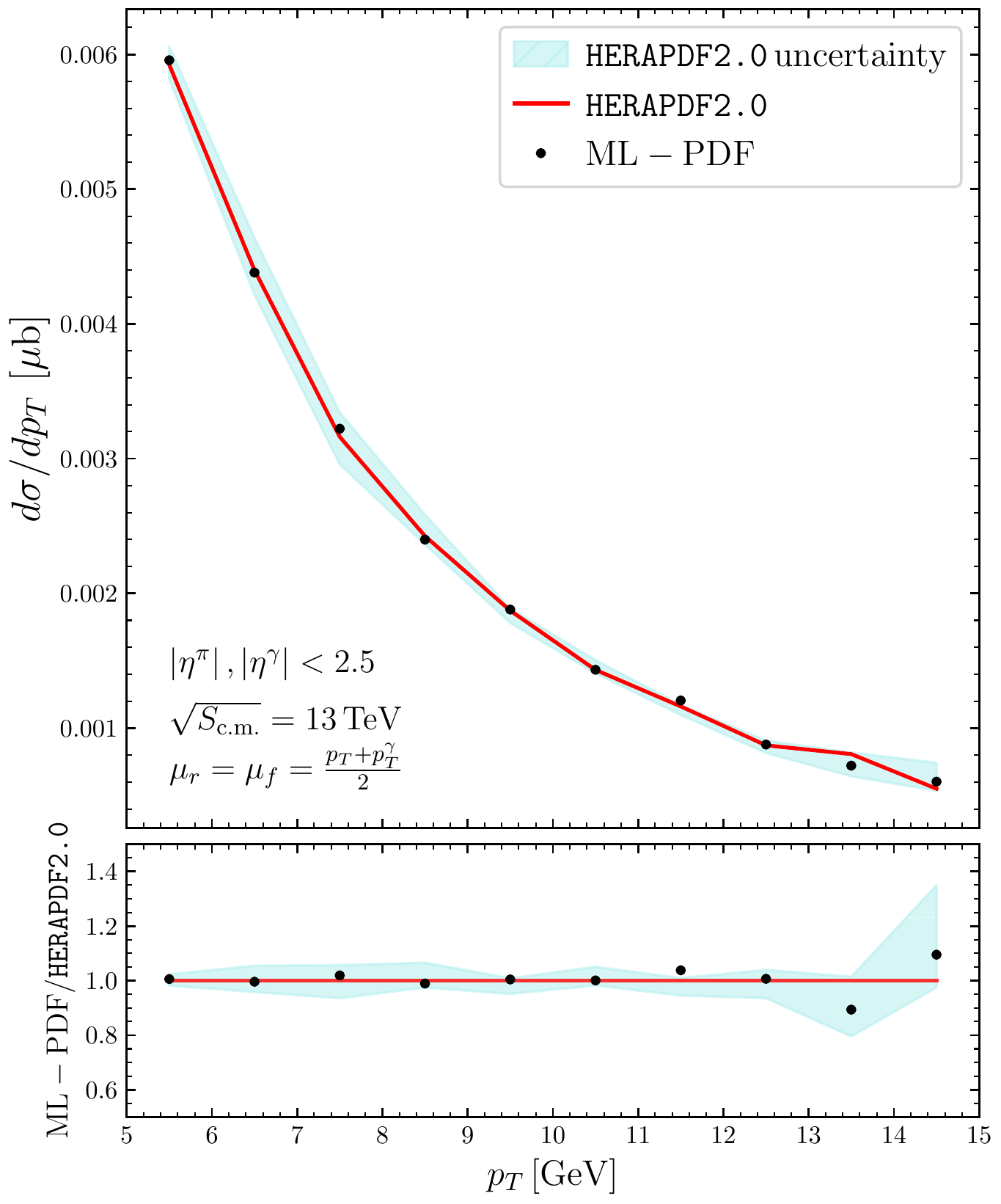}
\caption{Upper plot: differential cross-section for the process $p+p \to \gamma+\pi$ as a function of the transverse momentum of the pion ($p_T$), including up to NLO QCD corrections. We run the simulation using our analytic ML-PDFs (dots) and \texttt{LHAPDF} with \texttt{HERAPDF2.0} (solid). Lower plot: ratio between the two cross-sections.} 
\label{fig:final_gammapi}
\end{figure}

In the second example, we turned to a simulation that required a significantly larger amount of trials to reach a stable output. We tested our ML-PDFs with the calculation of $p+p\to \gamma+\pi$ including up to NLO QCD effects \cite{deFlorian:2010vy,Renteria-Estrada:2021rqp}, at $\sqrt{S_{\rm c.m.}}=13$ TeV. We consider the central rapidity region, $|\eta^\pi|,|\eta^\gamma|\leq 2.5$, imposing $p_T^\gamma \in [30 \, {\rm GeV},1500 \, {\rm GeV}]$. We used the same FFs as before, and we took the renormalization and factorization scales to be equal to
\beq
\mu = \frac{p_T+p_T^\gamma}{2} \, .
\eeq
In this case, we chose to run $10^8$ points both in order to be in what we expect (from Tab. \ref{tab:speed}) to be a very low time-gain region and because the observable requires a large number of points to produce physical results. We show the outcome in Fig. \ref{fig:final_gammapi}. As before, the upper plot presents the cross-section, while the lower plot demonstrates the ratio between the two computations. Given the $p_T$ range explored, there are some oscillations of the output associated to statistical fluctuations. In fact, using $10^7$ points gave unphysical results for some $p_T$ values. We can then say that this observable would greatly benefit from using a higher number of points. For our purpose, $10^8$ points is enough. In the $p_T \leq 10$ GeV region, that dominates the cross-section, the differences are within { $5\%$ and are feasible of further reduction.  The runtime for \texttt{LHAPDF} was of 207.5 minutes, almost 3.45 hours, while the ML-PDF run took 144.5 minutes, more than a full hour less.} We summarise the runtime of the cross-sections in Tab. \ref{tab:speed2}. This gain is much larger than in the previous test, and far more than expected from the table. {Also, we notice that the predictions obtained with our ML-PDFs are compatible with the propagation of errors of \texttt{HERAPDF2.0} into the observable (light blue band).} 

\begin{table}
\begin{center}
\begin{tabular}{|c |c |c |c|} 
 \hline
 Obs. & \texttt{LHAPDF} (s) & \, ML-PDFs (s) \,& \, Gain($\%$) \, \\ [0.5ex] 
 \hline\hline
 $p+p\to \pi$ & $628.320$ & $558.854$ &  \,$11.06$\\ 
 \hline
\, $p+p\to \gamma+\pi$ \, & \, $12452.273$ \, & \, $8671.827$  & \, $30.36$   \\  
 \hline
\end{tabular}
\caption{Comparison of the time (in seconds) required to compute two observables using \texttt{LHAPDF} and ML-PDFs. See text for details.}
\label{tab:speed2}
\end{center}
\end{table}

To conclude, the results of these two realistic calculations clearly supports the potential of our ML-PDFs to reduce the computational cost of the simulations, keeping under control the uncertainties of the approximation.

\section{Conclusions}
\label{sec:Conclusions}
{In this work, we provide an analytic approximation to PDFs using machine-assisted techniques to adjust both their $x$ and $Q$-dependence. Our starting hypothesis was the assumption that the $x$-dependence could be reproduced by Eulerian-like functions, whilst all the $Q$-dependence is embodied within the coefficients.}

By doing so, we obtain a reliable approximation to \texttt{HERAPDF2.0}, taking into consideration up to NLO QCD corrections. We show that the integral error is under control for an ample range of $x$ and $Q$ values. In fact, our ML-PDFs were tested for $Q \in [Q_0, 1000 \, {\rm GeV}]$, presenting (for most of the distributions) deviations w.r.t. \texttt{HERAPDF2.0} below percent level. It is important to highlight that this is comparable with the error of the PDF sets themselves, and far smaller than the uncertainties introduced by truncating the perturbative expansion of observables in QCD. In fact, we find that the error induced by the ML-PDF in $p+p \to \pi$ 
is ${\cal O}(1 \, \%)$ whilst for $p+p \to \gamma+\pi$ is ${\cal O}(5 \, \%)$ \footnote{In this work, we neglect to study the uncertainty in the parameters of the ML-PDFs. Such a study is beyond the scope of the present work, and we leave it to future studies.}. These errors are much smaller than the ${\cal O}(20-50 \, \%)$ -or even larger- uncertainties arising from scale variations. Furthermore, {we found a non-negligible reduction of the runtime: ${\cal O}(11 \, \%)$ for $p+p \to \pi$ and ${\cal O}(30 \, \%)$ for $p+p \to \gamma + \pi$.}

{To conclude, we want to emphasize that the strategy explained in this article is fully applicable to any PDF set, and even to FFs, at any perturbative order (NLO, NNLO and beyond), due to the fact that we made no assumptions about the order at which DGLAP evolution is truncated. The only required ingredient is a previously existent PDF, extracted from data, to be used as input in our formalism.} Our only hypothesis is that the $Q$-dependence is fully embodied within the coefficients $\{A,B,\ldots\}$ described in Sec. \ref{sec:Results}. Following this approach, {we can avoid using interpolation methods to evaluate the distributions} and speed up the calculation of PDFs (and, eventually, FFs), {because we obtain a fit that already embodied the evolution in an analytically approximated way}. This does not only bring a reduction of the CPU time, but also leads to a reduction of the CO$_{\rm 2}$ footprint and points towards more efficient and sustainable practises in high-energy physics (HEP) \cite{Banerjee:2023avd}. For this reason, further developments to obtain analytic approximations to PDFs/FFs are expected and this work can be considered as a first proof-of-concept paving the way for future improvements. As final remark, we want to stress that our approach does not replace the PDF extractions from data; in fact, they constitute an essential input and the basis for ML-PDF determinations as the one presented here.

\section*{Acknowledgements}
We would like to thank L. Cieri, V. Mateu and G. Rodrigo for fruitful comments about the manuscript. This work is supported by the Spanish Government (Agencia Estatal de Investigaci\'on MCIN /AEI/10.13039/501100011033) Grants No. PID2020-114473GB-I00, PID2022-141910NB-I00 and Generalitat Valenciana Grants No. PROMETEO/2021/071 and ASFAE/2022/009. The work of D. F. R.-E. is supported by Generalitat Valenciana (CIGRIS/2022/145).  S. A. O.-O and R. J. H.-P. are funded by CONAHCyT through Project No.~320856 (Paradigmas y Controversias de la Ciencia 2022) and Ciencia de Frontera 2021-2042. R. J. H.-P. is also funded by Sistema Nacional de Investigadores from CONAHCyT. P.Z. acknowledges support, at the early stages of this work, from the Deutsche Forschungsgemeinschaft (DFG, German Research Foundation) - Research Unit FOR 2926, grant number 430915485. P.Z. is funded by the ``Atracción de Talento'' Investigador program of the Comunidad de Madrid (Spain) No. 2022-T1/TIC-24024. The work of G.S. was partially supported by EU Horizon 2020 research and innovation program STRONG-2020 project under grant agreement No. 824093 and H2020-MSCA-COFUND USAL4EXCELLENCE-PROOPI-391 project under grant agreement No 101034371.



%

\end{document}